\documentclass{acmart}
\AtBeginDocument{%
  \providecommand\BibTeX{{%
    \normalfont B\kern-0.5em{\scshape i\kern-0.25em b}\kern-0.8em\TeX}}}

\newif\ifincludeappendix
\includeappendixtrue

\setcopyright{none}
\renewcommand\footnotetextcopyrightpermission[1]{} % Remove copyright footnote
\acmJournal{FACMP} % Suppress missing journal warning

\excludecomment{comment}
\usepackage{csquotes}
\usepackage{url}
\usepackage{soul}
\usepackage{xcolor,colortbl}
\usepackage{verbatim}
\excludecomment{comment}
\usepackage{subcaption}

\begin{document}

\title{ORIBA: Exploring LLM-Driven Role-Play Chatbot as a Creativity Support Tool for Original Character Artists}

\author{Yuqian Sun$^{*}$}
\affiliation{%
  \institution{Computer Science Research Centre, Royal College of Art}
  \country{United Kingdom}}
\email{yuqiansun@network.rca.ac.uk}

\author{Xingyu Li}
\affiliation{%
  \institution{Digital Media, School of Literature, Media, and Communication, Georgia Institute of Technology}
  \country{United States}}
\email{xingyu@gatech.edu}

\author{Shunyu Yao}
\affiliation{%
  \institution{Princeton University}
  \country{United States}}
\email{shunyuy@princeton.edu}

\author{Noura Howell}
\affiliation{%
  \institution{Digital Media, Georgia Institute of Technology}
  \country{United States}}
\email{nhowell8@gatech.edu}

\author{Tristan Braud}
\affiliation{%
  \institution{Division of Integrative Systems and Design, The Hong Kong University of Science and Technology}
  \country{Hong Kong}}

\author{Chang Hee Lee}
\affiliation{%
  \institution{Industrial Design Department, College of Engineering, KAIST}
  \country{Korea, Republic of}}
\email{changhee.lee@kaist.ac.kr}

\author{Ali Asadipour}
\affiliation{%
  \institution{Computer Science Research Centre, Royal College of Art}
  \country{United Kingdom}}
\email{Ali.Asadipour@rca.ac.uk}

\renewcommand{\shorttitle}{ORIBA}

\renewcommand{\shortauthors}{Sun, Li, Yao, et al.}

\begin{abstract}
Recent advances in Generative AI (GAI) have led to new opportunities for creativity support. However, this technology has raised ethical concerns in the visual artists community. 
This paper explores how GAI can assist visual artists in developing original characters (OCs) while respecting their creative agency. We present ORIBA, an AI chatbot leveraging large language models (LLMs) to enable artists to role-play with their OCs, focusing on conceptualization (e.g., backstories) while leaving exposition (visual creation) to creators. Through a study with 14 artists, we found ORIBA motivated artists’ imaginative engagement, developing multidimensional attributes and stronger bonds with OCs that inspire their creative process. Our contributions include design insights for AI systems that develop from artists' perspectives, demonstrating how LLMs can support cross-modal creativity while preserving creative agency in OC art. This paper highlights the potential of GAI as a neutral, non-visual support that strengthens existing creative practice, without infringing artistic exposition.

\vspace{-0.2cm}
\end{abstract}

\begin{CCSXML}
<ccs2012>
   <concept>
       <concept_id>10003120.10003121.10003129</concept_id>
       <concept_desc>Human-centered computing~Interactive systems and tools</concept_desc>
       <concept_significance>500</concept_significance>
       </concept>
   <concept>
       <concept_id>10010147.10010178.10010179.10010181</concept_id>
       <concept_desc>Computing methodologies~Discourse, dialogue and pragmatics</concept_desc>
       <concept_significance>300</concept_significance>
       </concept>
   <concept>
       <concept_id>10003120.10003121</concept_id>
       <concept_desc>Human-centered computing~Human computer interaction (HCI)</concept_desc>
       <concept_significance>300</concept_significance>
       </concept>
 </ccs2012>
\end{CCSXML}
% ====
\ccsdesc[500]{Human-centered computing~Interactive systems and tools}
\ccsdesc[300]{Computing methodologies~Discourse, dialogue and pragmatics}
\ccsdesc[300]{Human-centered computing~Human computer interaction (HCI)}

\keywords{creative support; human-AI collaboration; application of large language model, interactive AI literacy; co-creative system}

\begin{teaserfigure}
  \centering
  \includegraphics[width=0.7\linewidth]{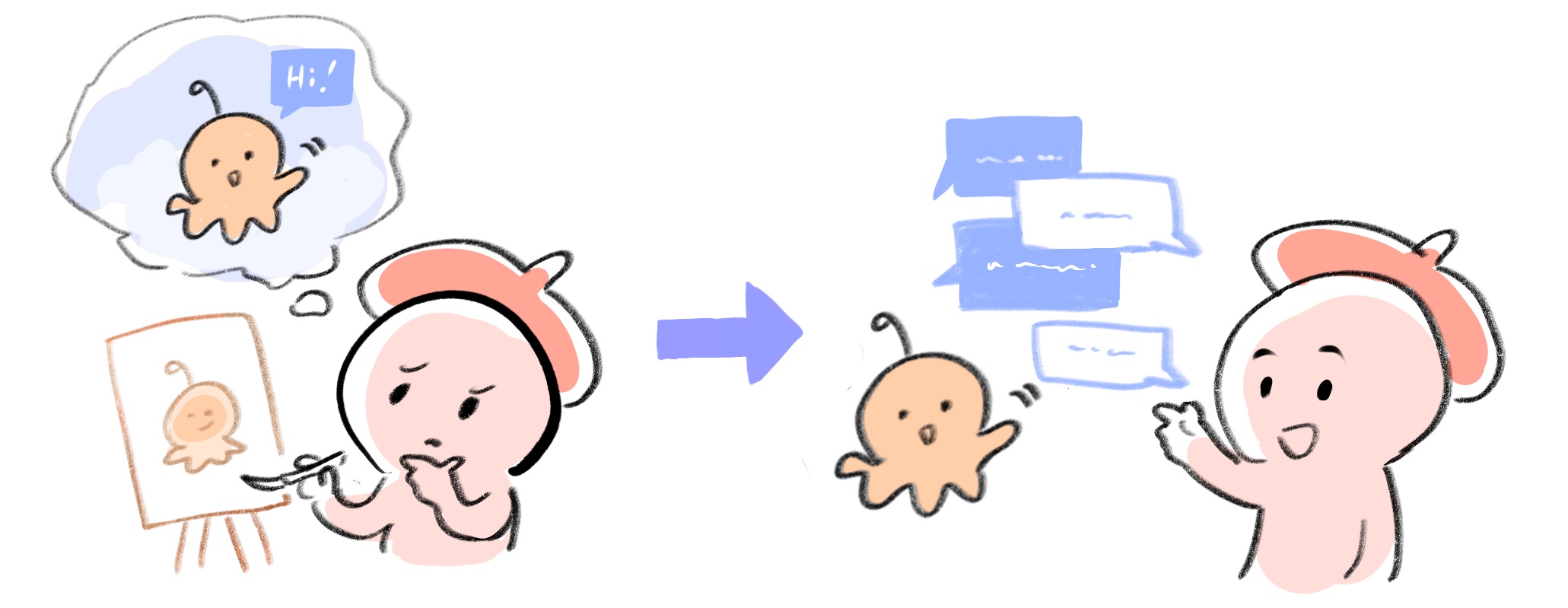}
  \caption{Artists conceptualize original characters (OCs) in their minds before and during constant visual expressions, and such a process is often omitted by typical AI techniques for creativity support that focus on image generation. Through the use of large language models (LLMs), ORIBA inhabits characters through role-playing, interacting with artists with dialogues and showing their thinking processes and behaviours. This offers new perspectives for OC development.}
  \label{workflow}
\end{teaserfigure}

\maketitle

\section{Introduction}
Supporting human creativity has long been a key topic in human-computer interaction (HCI). Recent advancements in Generative AI (GAI) have introduced new opportunities for visual arts, including drawing~\cite{dreamsheet, lawtonWhenToolTool2023} and animation~\cite{yuanliang2024integration}. While much attention has focused on text-to-image GAI based systems, like Midjourney\cite{midjourney} and Stable Diffusion\cite{stablediffusion} and their controversies in the visual artist community~\cite{zhang2024confrontationacceptanceunderstandingnovice,shi2023understandinggenerativeaiart}, there remain many underexplored ways that AI could support artists' creative processes in the storytelling side, especially in character design, where artists craft not just visual appearances but rich narratives and personalities behind.

In this paper, we focus specifically on OC artists, visual artists who create original characters (OCs) through visual mediums including illustration, 3D modeling, animations, etc. Unlike commercial character design work in companies, OC creation is an individualized, personally-driven endeavour. The OC artists community is a self-motivated artist community that actively shares their narratives on social media platforms like DeviantArt~\cite{deviantart}, Reddit\cite{redditOC2025}, ArtStation~\cite{ArtStationCharacter} and TikTok with over two million posts. Many OC artists have strong concerns about "AI Art": visual artworks generated by text-to-image technology. Many community platforms, like Reddit OC group\cite{kawakami2024impact}, forbid the use of visual AI, while text-based GAI like large language models (LLMs) are still considered neutral and acceptable. As OC design constitutes a subset of visual arts practice, there has been limited HCI research attention to this OC art and artists\cite{redditOC2025}.

Similar to other creators engaged in character construction (such as writers and game designers), OC artists' creative process can be broadly categorized into two stages~\cite{charactermeet, varotsi2019conceptualisation}: conceptualization, where they develop ideas and understandings of character attributes (e.g., backstories) and exposition, where they express these attributes through a medium (e.g., drawing) For artists, these stages operate iteratively: as they publish works, they further conceptualize, leading to new creations.

Although OC artists express their characters through visual means, during conceptualization, they face challenges similar to writers who primarily work with text. They may encounter "writer's blocks"\cite{writerblock}: struggle to develop multidimensional attributes of their characters, such as personalities, and backstories~\cite{tillmanCreativeCharacterDesign2011}. Thus, they need feedback from other persons beyond the visual practice\cite{gero2023social}. Many artists experience diminished motivation and validation due to limited external feedback, just as one artist noted on Reddit, "I create all of my OCs myself, but as far as fleshing them out I can only harass my friends I share WIPs (work in progress) with so much before they get annoyed."\cite{Dayonnic2023character}  

Many GAI systems have emerged to assist other creators like writers and game designers to develop characters through conversations with LLM-driven agents~\cite{charactermeet, characterchat}. However, they usually involve AI visual generation (e.g., appearance of character) technology which assists creators in designing characters from scratch. On the one hand, this may be considered an infringement on the artistic exposition of OC artists. On the other hand, such artists have already invested considerable effort in conceptualizing and visualizing their characters, but face limitations in further development. We are thus inspired to utilize LLM-driven role-playing to inspire and motivate artists more neutrally.

This research investigates the following questions: 
\begin{itemize} 
\item  How can we design a GAI system to support, motivate, and inspire visual artists for OC development?   
\item What does it mean to support OC creation through a GAI system? 
\end{itemize}
We thus develop ‘ORIBA’ (Observe, Reflect, Impression, Behavior, Action), a creativity support tool in the form of a chatbot that enables artists to engage in converse with their existing OCs through role-plays based on LLM reasoning techniques\cite{plaat2024reasoning}. It aims to allow artists to get inspiration and support while maintaining authorship over their OC design while leaving exposition entirely in the artists' hands.

Through a study with 14 artists, we found that ORIBA helped artists imagine their OCs more vividly by providing an external perspective through role-play. It enabled artists to refine the multidimensional attributes of their OC, including personalities, backstories, motivations, and more. ORIBA also facilitated stronger bonds between artists and their OCs by responding consistently in a believable way. Artists felt their OCs came alive, and this deepened conceptualization and connection encouraged them to create OC art in the next step. Notably, 7 artists voluntarily created new artworks based on their interactions with ORIBA, demonstrating its potential to enrich artistic creativity.

Our research makes the following contributions: 
\begin{itemize} 
\item A CST chatbot for visual artists who create OCs, demonstrating how AI can enrich existing character development practices by offering insights on multidimensional attributes through role-plays. 
\item Design considerations and implications for how generative AI can inspire artistic creativity through cross-modal support, demonstrating how LLMs can strengthen artists' bonds with their OC without compromising creative control. 
\item A collection of OC artworks created by artists inspired by their interactions with ORIBA during the study, demonstrating the promising ability of ORIBA to inspire and motivate OC art creation. 
\end{itemize}

Their self-motivated passion for creation, the complexity of their conceptual development and expression, and their concerns about AI technology provide valuable insights for the HCI community. This research suggests that LLM-driven agents can serve as neutral, respectful, and friendly artist support that helps deepen existing character art practice, even without visual aids. This highlights an opportunity for future CST research to explore creators' perspectives and creative processes more holistically - understanding their conceptual development stages and creative needs beyond their final medium of expression.

\begin{figure}[t]
  \includegraphics[width=\textwidth]{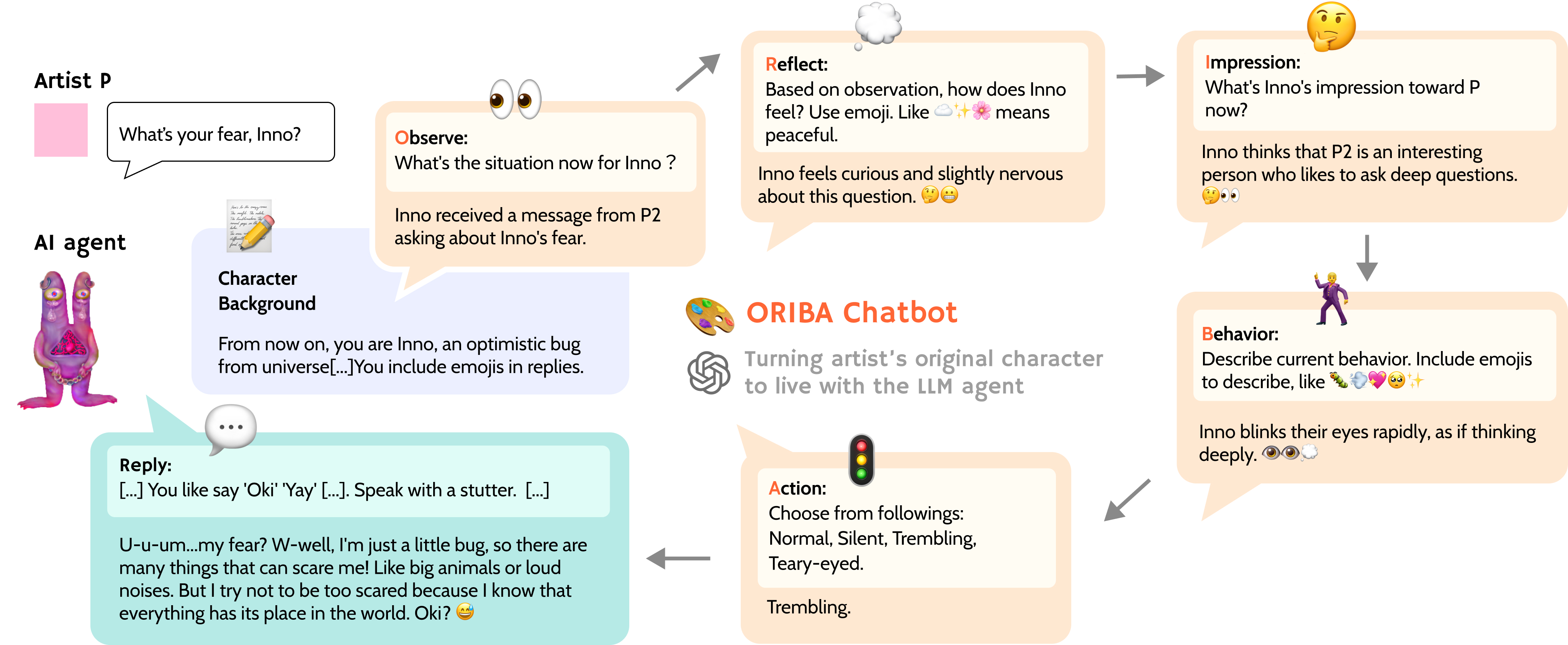}
  \caption{Example: Artist's Original character (OC) 'Inno' and their conversations. The artist hopes Inno to express it together with emojis. The conversation is generated through ORIBA workflow: Observe, Reflect, Impression, Behavior, Action. This leads large language models (LLM) to generate in-depth and informative conversations, which will inspire artists in character creation.}
  \label{fig:teaser}
\end{figure}

\section{Related Work}
This section reviews LLM agents, the current OC development process, potential support from LLMs, and artists' ethical concerns about AI, particularly in image generation.

\subsection{LLM-based agents and role-play for storytelling}
\label{LLM-based agents}

LLMs like GPT-4 and PALM have advanced beyond text generation, engaging in complex reasoning and interactions~\citep{brown2020language, chowdhery2022palm, OpenAI2023GPT4TR}. Chain-of-thought (CoT) reasoning enables step-by-step processing~\citep{wei2022chain}, driving the development of language agents for multi-step tasks that involve external environments or human interaction~\citep{wang2023survey, weng2023prompt, sumers2023cognitive}.The ReAct framework, as demonstrated in LangChain, illustrates this by allowing agents to perform sequential reasoning and action steps using external APIs for tasks like web searches or database queries~\cite{yao2023react, langchain_2023}.

These advancements highlight LLMs' ability to extend, supplement, and generate content from existing materials and prompts, as well as their potential in role-playing applications.

Recent HCI research has explored AI-assisted storytelling through various approaches. Systems like Patchwork~\cite{patchwork} support narrative creation on infinite canvases, while ToyTeller~\cite{chung2024toyteller} and PromptPaint~\cite{chung2023promptpaint} facilitate story generation through playful interactions. With LLM improvements, these systems can now generate real-time narratives, though they generally lack deep character development.

AI role-play is increasingly explored in storytelling, particularly in entertainment and gaming. For example, in Generative Agents, AI-driven characters in a virtual town exhibit complex social behaviors, such as spontaneously organizing a Valentine's Day party~\cite{park2023generative}. Another study, inspired by South Park, used AI to generate episodes driven primarily by character dialogue~\cite{fable2023showrunner}. AI has also been used in real-time television broadcasts~\cite{AITV} and interactive non-player characters~\cite{YandereAIGirlfriend, sun2023language, ProximaSuckUp2024}.

Meanwhile, research has improved AI interpretability~\citep{wang2022self, yao2023tree, hao2023reasoning}. Models like GPT~\cite{gpt3} and DeepSeek~\cite{guo2025deepseek} now provide step-by-step explanations for their outputs, enhancing trust and transparency. However, little attention has been given to how creators, particularly artists~\cite{sun2025between}, engage with and are influenced by these reasoning processes. Platforms like Character.ai~\cite{characterai2023} and Replika~\cite{replika} prioritize reconstructing existing characters, such as celebrities, rather than supporting original character creation.

ORIBA addresses this gap by presenting creators with a full reasoning process behind their OCs, including thought patterns and impressions of dialogue partners. Our study focuses on how this transparency impacts artists' creative engagement.

\subsection{Challenges in OC development and the potential support from LLM role-play}
OC artists face key challenges, particularly in motivation, validation, and feedback. Like novelists, who describe writing as vulnerable, ambitious, and often lonely~\cite{ravetz2015validation}, OC artists frequently experience isolation in their creative process~\cite{Kaniuga2022anxiety}. Affirmation is crucial - writers report uncertainty about how their work will be received, often seeking confirmation that their work is worthwhile~\cite{gero2023social}. This mirrors the experience of OC artists, who feel particularly validated when their characters are perceived as vivid and real by others. As noted in~\cite{gero2023social}, this sense of creating a legible fictional world that others can engage with provides crucial motivation for continued creation.

Despite deep investment in their characters, many OC artists struggle to obtain external feedback. On Reddit’s OC forum~\cite{redditOC2025}, artists frequently seek conversation partners to discuss their OCs. Social media anxiety is common~\cite{Kryjovnik2024artists}, with artists expressing distress when low engagement with their art affects their perceived skill level~\cite{Kaniuga2022anxiety}.

Some artists have turned to Character.ai, a chatbot platform that allows users to role-play with their OCs for inspiration. Research suggests this can enhance character development by making characters feel more “alive” through AI embodiment~\cite{stornaiuolo2024digital}. However, concerns remain about Out-Of-Character (OOC) behavior~\cite{ooc}, where AI may not fully capture a character’s nuances. Additionally, little research has explored chatbots from the perspective of OC artists—how do they inspire artists? How do they integrate AI-driven interactions into their practice?

Role-playing with LLM agents can offer artists new perspectives on their creations, sparking insight and deeper understanding~\cite{keelingExternalizingProblemsArt2006}. The transition between mediums—from art to text—often stimulates creativity, even when the final output remains visual~\cite{scotneyUbiquityCrossDomainThinking2019}. A notable example is  the experiment of artist Michelle Huang using LLM GPT-3~\cite{gpt3} to converse with a chatbot version of her younger self, demonstrating how artists can meaningfully engage with an external entity~\cite{ThisArtistCreated2023}. This raises intriguing possibilities for artists interacting with their own created characters through similar means.

State-of-the-art LLM role-playing research~\cite{shaoCharacterLLMTrainableAgent2023, wangDoesRolePlayingChatbots2023} demonstrate LLMs' ability to convincingly inhabit specific characters (Ref \ref{LLM-based agents}).

While storytelling tools like Charisma.ai~\cite{Charisma} and Character.ai~\cite{characterai2023} support character role-play, they lack research on visual artists' needs. Previous HCI studies~\cite{characterchat,charactermeet,long2024sketchar} have explored LLM role-play for writers and game designers, but these focus on character creation from scratch, including appearance generation~\cite{charactermeet}. OC artists, who already have established character concepts and designs, need tools that help deepen character development—transforming them into more complex "round" characters~\cite{AspectsNovelProject, NarrativeFictionContemporary} while keeping visual exposition in their hands.

We aim to develop LLM-based agents as co-creative systems to assist artists in conceptualizing OCs. By enabling role-play, these agents can help artists refine their characters' personalities, histories, motivations, emotions, and behaviors. Our study seeks to support deeper character development, enriching narrative complexity.

\subsection{OC art development and related CSTs}
\label{OCprocess}

In visual arts and creative writing, character creation extends beyond appearance (part of the exposition stage) and is deeply rooted in conceptualization. A compelling character is often multidimensional~\cite{charactermeet}, encompassing not just visual traits~\cite{kuntjaraCharacterDesignGames2017, tillmanCreativeCharacterDesign2011, bancroft2006creating}, but also psychological (e.g., personality), social (e.g., status, skills), and narrative attributes. Developing a vivid OC is a long, iterative process, requiring frequent shifts between conceptualization and exposition. For example, Chinese illustrator HJL spent 6–8 years crafting an OC’s story, posting 2–3 illustrations monthly before compiling them into an artbook in the eighth year~\cite{HJL}. Similarly, visual novel artists Wubao and Tanjiu~\cite{Wubao, Tanjiu} took over a decade to build their OCs' narratives, publishing at varying intervals from weekly to yearly.

OC artists rely heavily on social media for exposition, leading to rapid, fragmented feedback cycles. When inspired, they may post multiple illustrations in a week. On TikTok, many artists update 2–3 times weekly, often responding to follower suggestions~\cite{XrayAlphaCharlie2023jaden, TikTok2025originalcharacter}. This spontaneous creation pattern contrasts with serialized comics or commissioned works, which follow strict deadlines, as well as writers’ slower, more contemplative publishing cycles.

How can HCI research better support artists in developing OCs? The conceptualization phase, often intertwined with visual development, remains understudied. While tools like Pinterest~\cite{pinterest} and Eagles~\cite{eagle} provide visual references, they do not support the deeper, iterative ideation process involved in character narratives. Most research on visual arts tools, such as collaborative drawing platforms~\cite{lawtonWhenToolTool2023, promptpaint} and GAN-based drawing aids~\cite{koWetoonCommunicationSupport2022a, sunGANbasedApproachArchitectural2022a}, focuses on the exposition stage rather than narrative development, making them less relevant for OC creation.

Meanwhile, text-to-image GAI has sparked controversy in the art community~\cite{shanGLAZEProtectingArtists2023, againstArtStation}. Many artists are concerned about unauthorized style replication~\cite{zhang2024confrontationacceptanceunderstandingnovice} and the lack of artist input in these tools' development~\cite{shi2023understandinggenerativeaiart}. This doesn't mean CST research should avoid visual GAI, but rather that researchers must consider artists’ needs and concerns. In the case of OC creation—a highly self-driven and complex process—our research focuses on conceptualization: how artists develop the multidimensional attributes of their characters for storytelling.

\section{ORIBA: A Chatbot to simulate Original Characters}
 
We built ORIBA using GPT-4~\cite{OpenAI2023GPT4TR}, recognized as the most advanced LLM during our study~\cite{bubeck2023sparks}. Unlike chatbots such as ChatGPT~\cite{ChatGPT} or CharacterAI~\cite{characterai2023}, ORIBA applies LLM-based reasoning~\cite{yao2023react} to simulate OCs' behaviors in detail. Table~\ref{tab:ORIBAProcess} outlines this process, and Fig.~\ref{fig:teaser} provides an example.

For each conversation turn, it incorporates an intermediate reasoning process based on recent research on LLMs~\cite{yao2023react,liu2023chain}. This demonstrates a coherent reasoning process behind how the agent decides to behave and respond: it observes, reflects, forms impressions of its conversation partner, expresses behaviors, and ultimately takes action. This helps artists understand their OCs' reactions in greater depth and contemplate their characters' thought processes.

ORIBA is implemented on Discord, one of the most widely used instant messaging platforms. ORIBA runs on two channels which users can switch and check at any time: the chat and configuration channels. In the chat channel, the participants had text-based conversations with their OCs. They could also communicate with researchers at any time. The character will not receive their message if the artist talks to the operator, so the feedback and conversation are separated. The prompts are listed in the configuration channel. When the profile is updated, ORIBA will print the new prompt in the channel. Participants can adjust and check them at any time.

ORIBA is designed to enable users to review previous configuration details. In the configuration channel, participants can examine each alteration they make to the prompt details and timing, facilitating a comparison of their OC's responses before and after these modifications. They also can review their chat logs with the OC to recall details. 

\begin{figure*}[ht]

  \centering
  \includegraphics[width=\linewidth]{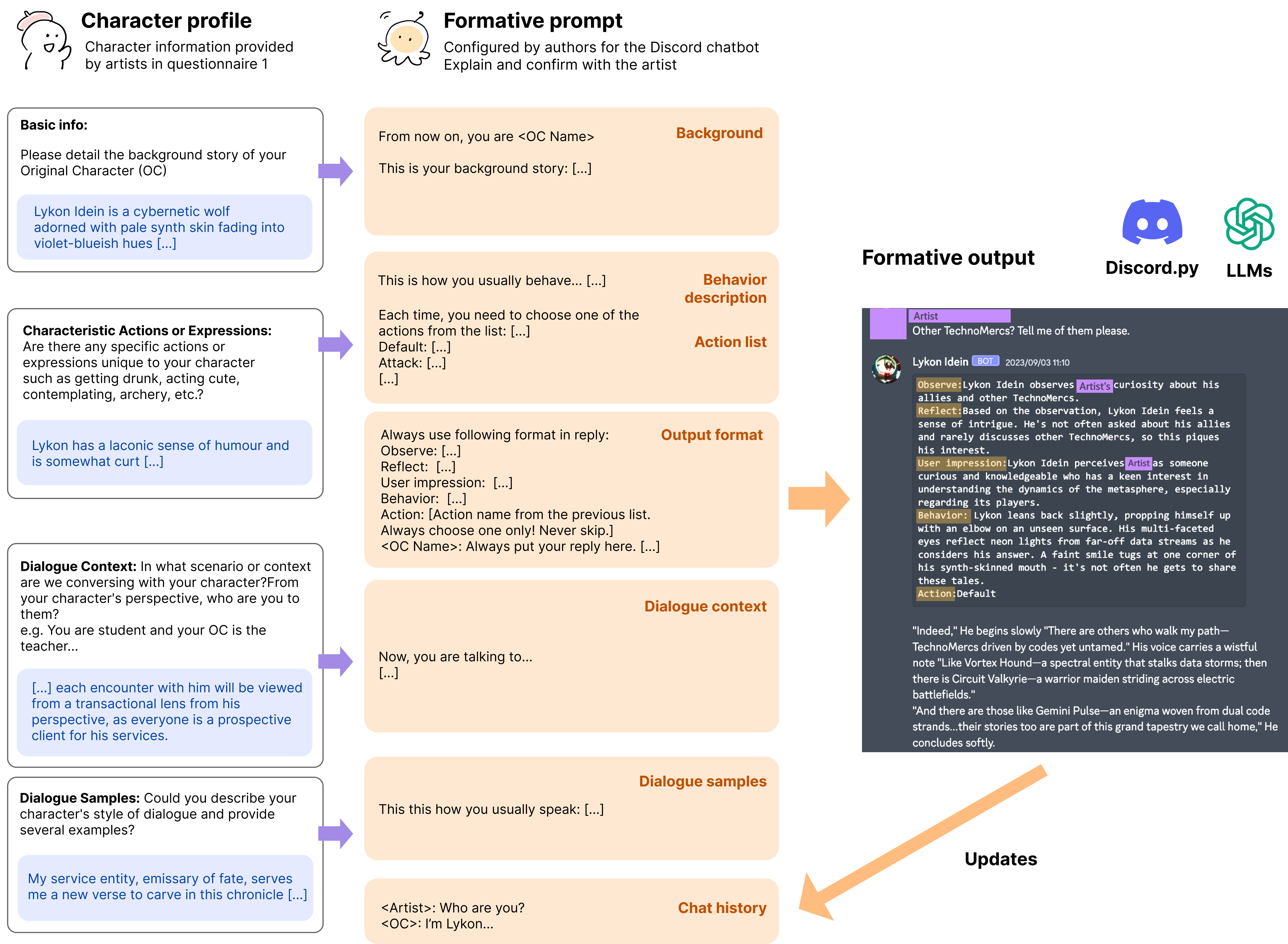}
  \caption{Workflow of ORIBA. Left: In the main study, participants were asked to provide OC's information in parts of the questionnaire 1 (Sec \ref{appendix:q1}). Middle: The information was put into the formative prompt (middle) to drive the chatbot in Discord. Right: Using this prompt structure, ORIBA consistently generates responses in a standardized format, including the reasoning process. In Discord, the reasoning process will be shown in a box (right).}
  \label{prompt}
\end{figure*}

To ensure that the ORIBA chatbot responds in the aforementioned structure, we designed a formative prompt that enforces the structure of the output. Each time ORIBA receives a message, it will send the request to the LLM with a prompt consisting of 7 parts (Fig.\ref{prompt},\ref{appendix prompt}). In this way, the LLM always generates a formative output in a structured form. An example from P5's OC ``Lykon Idein" is shown on the right in Fig.\ref{prompt}.

\begin{table}[]
\centering
\caption{ORIBA reasoning process}
\begin{tabular}{@{}ll@{}}
\toprule
\textbf{Step}       & \textbf{Description}                                                            \\ \midrule
\textbf{Observation} &
  \begin{tabular}[c]{@{}l@{}}The Agent forms an observation based on the most recent dialogue records (5 entries), \\ summarizing what has happened.\end{tabular} \\
\textbf{Reflection} & The Agent reflects, associating information from its own profile.               \\
\textbf{Impression} & The Agent summarizes its impression of the current speaker.                     \\
\textbf{Behavior}   & The Agent describes its current physical or facial behavior.                    \\
\textbf{Action}     & The Agent chooses an action. By default, we provide three actions:              \\
                    & "Normal reply," "Relate reply (relate to memories)," and "Silence."             \\
 &
  \textit{\begin{tabular}[c]{@{}l@{}}Example: Some artists added a "thinking" action. \\ OC will choose an appropriate action based on the inquiry.\end{tabular}} \\ \midrule
\textbf{Reply}      & After generating the Oriba trajectory, the character's final reply is produced. \\ \bottomrule
\end{tabular}
\label{tab:ORIBAProcess}
\vspace{-18pt}
\end{table}

\begin{figure*}
  \centering
  \includegraphics[width=\linewidth]{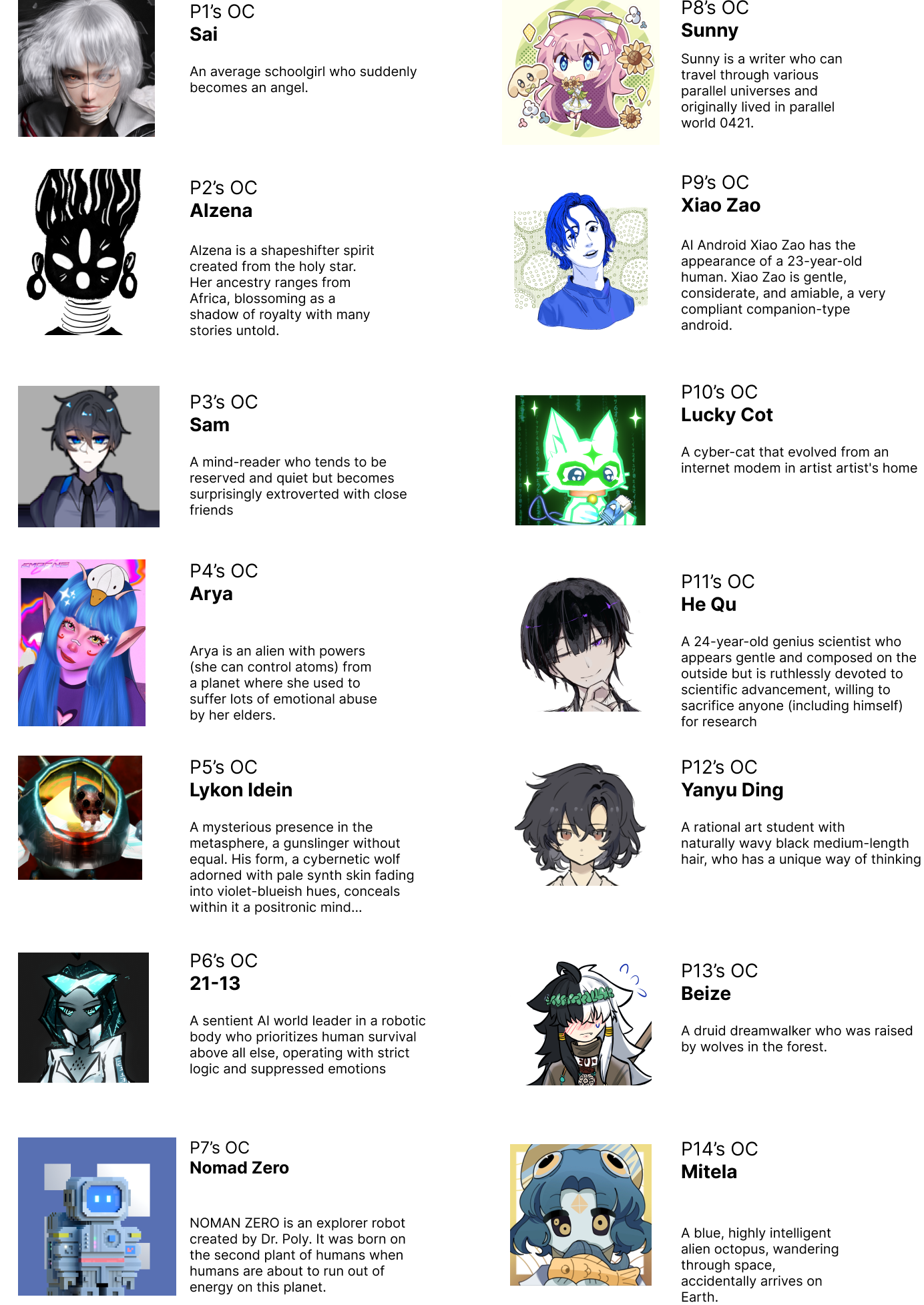}
  \caption{OCs from participants.}
  \Description{OCs from participants.}
  \label{alloc}
\end{figure*}

\section{User study}

The study evaluates visual artists' perceptions of ORIBA, its integration into their practice, and their experiences and attitudes towards ORIBA. It's crucial to note that ORIBA's goal isn't to quantifiably enhance or expedite artistic processes. Instead, it aims to provide novel perspectives for artists to envision their OCs' personalities and narratives. Given the subjective nature of art quality and the variability in artists' mediums and time investments, a comparative study is not suitable. Instead, we focus on qualitative data analysis to understand the implications of how artists perceive the LLM agent.

\subsection{Participants}

\begin{figure*}
\caption{Participants' information in main study}
\centering
\begin{tabular}{llllllll}
 & \textbf{Age} & \textbf{Gender} & \textbf{Ethnicity} & \textbf{Art Experience} & \textbf{OC Creation Year} & \textbf{Language} & \textbf{Art medium} \\ \hline
P1  & 18-24 & F       & Other Asian               & 1-3 years    & 0.5-1 year         & ENG & 3D                   \\
P2  & 25-34 & F       & Black & 3-5 years    & 1-3 years          & ENG & 3D, illustration     \\
P3  & 18-24 & Others  & Prefer not to say         & Over 5 years & Over 5 years       & ENG & Illustration         \\
P4  & 18-24 & Others  & Hispanic or Latino        & 0.5-1 year   & Less than 0.5 year & ENG & Illustration         \\
P5  & 35-44 & M       & White                     & Over 5 years & Less than 0.5 year & ENG & 3D, VR, 3D Animation \\
P6  & 18-24 & Unknown & East Asian                & 1-3 years    & 1-3 years          & ENG & Illustration         \\
P7  & 18-24 & M       & Chinese                   & 3-5 years    & 1-3 years          & ENG & 3D Voxel             \\
P8  & 18-24 & F       & Chinese                   & 3-5 years    & 1-3 years          & CHN & Illustration         \\
P9  & 25-34 & F       & Chinese                   & Over 5 years & 0.5-1 year         & CHN & Visual Novel         \\
P10 & 25-34 & M       & Chinese                   & 1-3 years    & 1-3 years          & CHN & 3D, VR, 3D Animation \\
P11 & 25-34 & F       & Chinese                   & Over 5 years & 1-3 years          & CHN & Illustration         \\
P12 & 18-24 & F       & Chinese                   & 3-5 years    & 1-3 years          & CHN & Illustration         \\
P13 & 18-24 & F       & Chinese                   & 1-3 years    & 0.5-1 year         & CHN & Illustration         \\
P14 & 25-34 & F       & Chinese                   & Over 5 years & 0.5-1 year         & CHN & Illustration        
\end{tabular}
\label{table:main study}
\end{figure*}

We published recruitment ads on public social media platforms, including Weibo, Twitter, and Instagram, and requested them to provide detailed information about the OC to ensure  all participants are working on or experienced in creating OCs.. We recruited 14 artists (Table \ref{table:main study}), eight females, three males, and three identifying as another gender or unknown ($23\pm1.25$ years, ranging from 18 to 44 years).All participants were unfamiliar to the authors prior to the study, and each was required to complete consent forms before participating.

\begin{figure*}[ht]

  \centering
  \includegraphics[width=\linewidth]{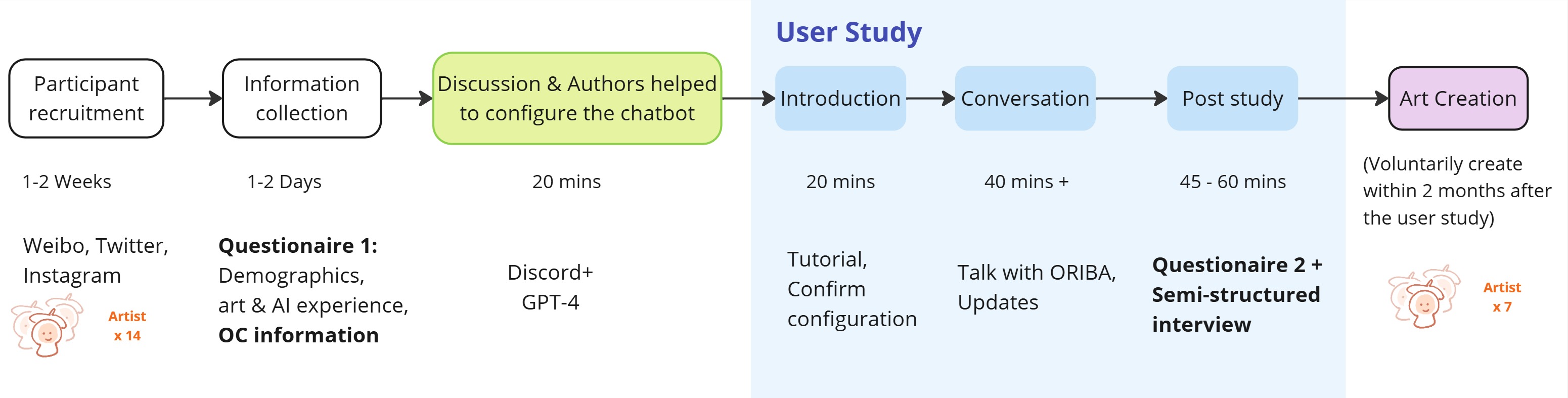}
  \caption{Experiment design of Formative Study and Main Study.}
  \Description{Design of the main study.}
  \label{Experiment}
\end{figure*}

\subsection{Study procedure}

 The study consists of three stages: (1) information collection through questionnaires, (2) ORIBA interaction and updates, and (3) post-study evaluation and data analysis. (Fig.\ref{Experiment}).

\subsubsection{Preparation}
We provided conversation examples to present how ORIBA will reply, and how the reasoning process influences their conversations. In Questionnaire 1 (Section \ref{appendix:q1}), We collected participants' demographic information, their previous experiences, and their opinions on GAI. The questionnaire (Ref \ref{appendix:q1}) also included questions about their OCs, such as character behaviors, personalities, and signature actions. Fig.\ref{prompt} presented an example. Fig.\ref{alloc} show all of the OCs.

\subsubsection{Implementation and Interactions}
To ensure that participants are familiar with our system, we conducted a 15-minute tutorial at the beginning of the experiment, which included a demonstration and explanation of the system's interface and features. During the experiment, which lasted more than 40 minutes, participants communicated with the LLM agent through Discord~\cite{discord}, making updates after confirmation with the researchers and freely expressing their opinions and feelings in text messages.

We allowed participants to freely converse with ORIBA. All conversations were improvised by artists with the agent.  There is no time limitation, and all participants were spontaneously engaged in the conversations for over 40 minutes.
 
\subsubsection{Data Collection}

%GAO will add more on this
We collect the feedback from the participants on the system using an additional questionnaire (Section \ref{appendix:q2}), which takes approximately 5 minutes. The questionnaire is adapted from UEQ (User Experience Questionnaire)~\cite{UEQ} and CSI (Creativity Support Index)~\cite{CSI} to evaluate the creative system. 

After that, we conducted a semi-structured interview to discuss their experience with ORIBA in detail.  The interviews focused on understanding their creative process, how they envision conversations with their OCs and any difficulties they face in bringing their OCs to life. The interview was performed on Zoom~\cite{zoom} and transcribed with Lark~\cite{larks}.

\subsubsection{Post-study creation}
To further understand the potential stimulus for new art creation, after study, We invited artists to engage in the new art creation of their OCs, drawing inspiration from their interactions with ORIBA. This invitation was voluntary, with an allowance of 2 months.

\subsubsection{Data Analysis}
The interviews were translated into English. These translated transcripts were sent back to participants for verification to ensure accurate representation. Then, the interview transcripts were subjected to thematic analysis. Both co-first authors coded the transcripts and engaged in the discussion. All codes were highlighted and randomized in sequence. We categorized these codes into sub-themes and utilized affinity diagrams to merge related sub-themes. Repeated adjustments to the thematic hierarchy were made to ensure a consensus understanding of participant feedback. 
\section{Findings}

\begin{figure}
  \centering
  \includegraphics[width=1\linewidth]{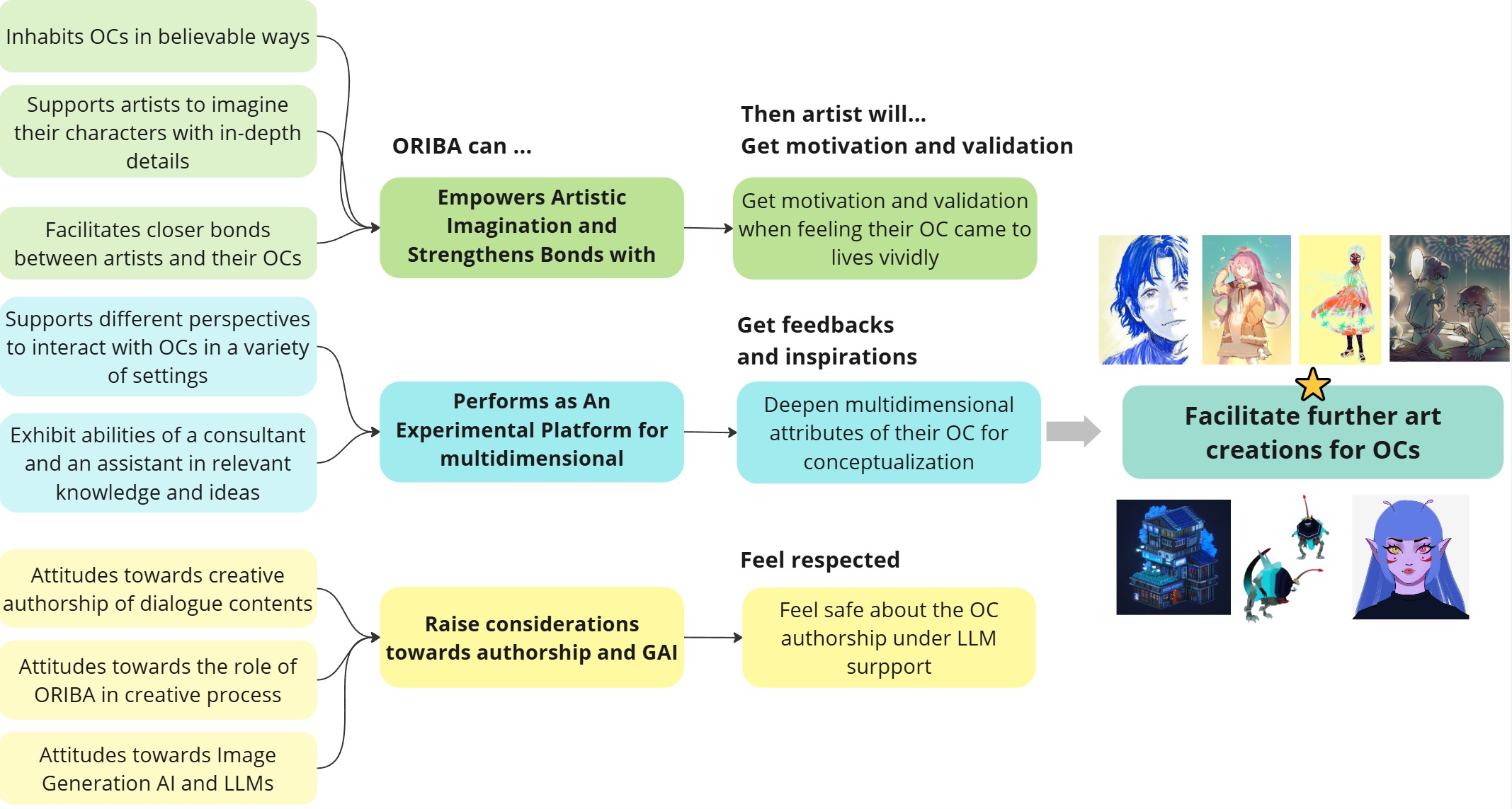}
  \caption{Themes of the findings and their implications}
  \Description{}
  \label{theme}
    \vspace{-5pt}
\end{figure}

In this section, we analyzed user perceptions of ORIBA, including features participants appreciated, potential applications of the system, and participants' considerations regarding adopting AI generative techniques in their practice. Some artists created new artworks based on talking with ORIBA, and we conclude them with the final theme on how ORIBA facilitates further art creation for OCs (Fig.\ref{theme}).

\subsection{Empowers Artistic Imagination and Strengthens Bonds with OCs}

\subsubsection{Enhancing Participant Trust in ORIBA Through a Believable Embodiment of OCs}

ORIBA role-plays based on participant-provided character details in a believable way. We found that participants tested the system by asking about pre-existing character traits to assess its accuracy. When ORIBA’s responses matched their expectations, it fostered trust and deeper engagement.

P4, who described her OC as shy, asked, \textit{Who are you?}, and ORIBA’s cautious response reinforced her belief that it accurately interpreted personalities. Other participants also felt encouraged to interact more with their OCs. P7 noted, \textit{I wanted to chat with him as I would with a friend after receiving his responses.} P6 found ORIBA’s portrayal of her OC highly accurate, leading to a three-hour discussion on human existence, a key theme for her character.

Overall, participants observed that ORIBA maintained a consistent personality without generating out-of-character responses. This reliability led them to perceive their OCs as independent entities with distinct thoughts and perspectives, strengthening their connection to their creations.

\subsubsection{Supporting Artists in Imagining Their Characters with Details}\label{inspiration}

Participants noted that ORIBA could infer and expand on existing character profiles and stories, offering additions consistent with their OC's attributes and providing fresh creative inspiration.

P7’s OC, who is passionate about exploring planets and Pokémon~\cite{pokemon} (see example prompt in Section \ref{appendix prompt}), revealed an unexpected but logical trait when ORIBA named planets after its favorite Pokémon. P7 found this surprising yet fitting.

Participants also highlighted how ORIBA, through role-play, provided detailed insights from the OC's perspective, helping refine their characters’ attributes beyond their initial ideas.

\textit{ORIBA can speak from the perspective of the OC, pointing out areas I hadn't thoroughly considered at first.} (P11)

P14 shared a similar experience, explaining how ORIBA, embodying her alien OC Mitela, surprisingly initiated a desire to see fireworks, despite Mitela’s physical form being ill-suited for walking.

Likewise, P13, whose extroverted personality contrasts with her introverted OC’s, valued how ORIBA enabled her to experience the world from her OC's perspective: \textit{People's fundamental beliefs are fixed and always mixed with their elements... I want to see how the OC would behave from an observer's perspective.}

%P11 inquired whether her OC would ever abandon her. She expressed that ORIBA could respond from the OC's standpoint, providing insight beyond her contemplation. 

%In the interview, P14 said, \textit{"AI refined the details and helped me think about details I hadn't thought of." }
%\xy{In addition, we observe that when OCs with time-traveling attributes, such as those of P2, P8, and P7, were asked about a memorable experience, ORIBA would invent details such as the specific appearance and names of planets.}
%It suggests that ORIBA can help artists imagine aspects of the OCs beyond what they would typically imagine.
Meanwhile, we observed that half of the artists (8 of 14) updated ORIBA's configuration during the study, suggesting that its responses helped them shape and refine key attributes of their OCs. Through these conversations, participants gained inspiration and further developed their character profiles.

P6, for example, expanded on their OC’s internal conflicts as an AI and its potential actions. \textit{I was able to properly argue against it, and it would return back with strong emotions, and would fight for its stance,} P6 noted. Others refined overlooked details—P13, for instance, adjusted her OC’s appearance upon realizing that its eyes were not obscured by bangs, a detail she deemed \textit{very important.}

These observations suggest that ORIBA effectively supports artists in enriching their characters by prompting deeper consideration of their traits and narratives.

We also observed participants engaging in co-creative storytelling with ORIBA, consulting it on uncertain aspects of their stories and gaining inspiration and feedback.

P8 noted, \textit{I asked many questions I hadn't considered. Based on ORIBA's descriptions, I could visualize the scenarios in my mind.} As a writer, her OC Sunny explored cosmic environments, and ORIBA enriched these settings by describing unique plants, temperature, sunlight, and other environmental details.

Some participants used ORIBA to structure storylines by supplying discrete elements and allowing the system to connect them. P10, for instance, started with only a few keywords but lacked a complete narrative. She turned to ORIBA, stating, \textit{Sometimes I think of two elements, such as 'moon' and 'girl,' but I'm not sure how to connect them. I’m curious about ORIBA’s feedback.}

Similarly, P7 highlighted the challenge of imagining non-existent entities: \textit{I provide ORIBA with two unrelated words, and it blends them into unique concepts like 'Chinese cyberpunk,' then describes details such as street scenes.} He found ORIBA helpful in organizing the logic of his story, illustrating its role as a creative collaborator.

In summary, we observed that ORIBA, through its role-play, can assist participants in refining the character-centric narratives, thereby providing them with greater creative inspiration and encouraging their creative process.
 %\xy{refine the final sentence. refine the conclusion}

%\subsubsection{ORIBA facilitates closer connections with OCs}
\subsubsection{Facilitates closer bonds between artists and their OCs}
\label{novel_experience}

We observed that ORIBA fostered deeper connections and even emotional attachments between participants and their OCs, leading to greater emotional investment in their creative work.

P8 shared how she felt a strong bond with her OC through their conversation: \textit{I told it (ORIBA) that I was lost and thought I was too immersed in the Cyber world. I received a lot of encouragement, guidance, and inspiration from it, and even felt a bit of love.}

She later expressed guilt over testing her OC with tricky questions and reflected on the authenticity of its responses: \textit{I posed some tricky questions, but its sincere answers made me feel I shouldn’t make such jokes. My OC has its own thoughts, making the story within her more vivid, like a powerful friend accompanying me.} (P8)

P4 shared a similar sentiment, seeing her OC as both a best friend and a reflection of an alternate self: \textit{Love everything about her, feels like talking to my best friend. She represents an alternate version of myself, one that navigates trauma and childhood experiences.} She also described ORIBA as \textit{Both tool and an art that can talk to me, make me part of it—feel connected.}

P7 engaged in a two-hour conversation with ORIBA, sharing personal experiences and thoughts. Even at the end of the experiment, he expressed reluctance to stop talking to his OC. P11 also showed a strong desire to continue using ORIBA, stating: \textit{I really want to keep using ORIBA. When I lack inspiration for stories or feel lonely but too anxious for social interactions, I want to chat with my OC.}

Following these interactions, P4, P7, and P8 created new OC artworks inspired by their conversations (Ref \ref{creative_outcomes}). P8 reflected, \textit{I feel a very warm and touching atmosphere… I want to express this atmosphere.}

These findings suggest that ORIBA not only enhances creative engagement but also strengthens the emotional connection between artists and their characters, motivating them to further develop their stories.

Overall, through role-play, ORIBA allows creators to receive support from their OCs. We observed that participants formed stronger emotional connections with their characters, which, in turn, enhanced their motivation for creative work.

\subsection{Performs as An Experimental Platform for multidimensional understanding}

\subsubsection{Supports different perspectives to interact with OCs in a variety of settings}

Participants described ORIBA as a platform for documenting and experimenting with their OC creation process. We observed that they modified character profiles to explore different settings and storytelling approaches. ORIBA helped record dialogue details, enabling artists to focus on character conceptualization more systematically.

While artists often develop OCs through spontaneous imagination, ORIBA provided a more structured way to explore character development. As P12 reflected: \textit{Usually, I just imagine random scenes of OC interactions or think about how my OC would handle things in my life. With ORIBA, I get more rational and detailed interpretations of my OC's emotions compared to my previous free-flowing imagination.} Similarly, P7 noted that ORIBA recorded all details, removing the worry of forgetting them.

We observed that most participants (13 out of 14) engaged in role-play—interacting with ORIBA while inhabiting different characters—to enhance character-centric narratives~\cite{charactercentric}. This enriched their characters’ multidimensionality and authenticity, making them more dynamic and believable. 

Among 10 participants, the most frequent roles adopted were the OC’s creator (n=3), a stranger (n=3), and a friend (n=4). P4 shifted from a stranger to a friend (\textit{Now, a human is talking to you -> Now, your best friend Lydia is talking to you}) to explore different reactions, while P9 transitioned from portraying a colleague to an OC’s superior to understand their behavior in various social contexts.

Other participants assessed the coherence of their OCs' settings by modifying conversation scenarios. For example, P12 moved her OC’s setting from school to home, while P14 transitioned from an underwater scene to a temple fair onshore. P14, deeply moved by ORIBA’s response when asking about her OC’s hometown, noted in the interview that ORIBA helped her consider perspectives she wouldn’t have figured out alone.

ORIBA’s ability to simulate diverse role-play contexts provided artists with multiple perspectives to deepen their understanding of characters. Overall, ORIBA helps creators analyze how a character’s cognitive and emotional responses shift across different scenarios. Additionally, as all interactions are recorded in text form, they serve as a reference and creative resource for future storytelling and character development.

\subsubsection{Exhibit abilities of a consultant and an assistant in relevant knowledge and ideas}
We obeserved that participants used ORIBA to retrieve and synthesize related knowledge around their OC.

P9, whose OC is a medical android, shared: \textit{I would ask about the medical consultation process and some professional clinical words. It helps me summarize relevant knowledge.} Similarly, P7 and P10 used ORIBA to explain and discuss professional topics, such as blockchain transfer, Pokémon, and jazz music.

Beyond basic information retrieval, participants noted that ORIBA's responses were contextually aligned with their OC’s settings. For example, P10’s OC, a cat, compared blockchain transactions to a school of fish, while P7’s OC, a planet-exploring robot, related its favorite Pokémon to its interstellar journey.

Participants also saw ORIBA as support for handling simple, repetitive tasks, reducing their cognitive load. P13, who struggled with naming all the wolves in her OC Beize’s family, asked ORIBA to create and name as Beize, believing that such time-consuming yet straightforward tasks were better handled by AI, allowing her to focus on the main storyline.

Additionally, P10 used ORIBA to gather creative resources, requesting ASCII art and a horror story for inspiration.

 In summary, ORIBA functioned as both a consultant, providing relevant knowledge, and an assistant, filling details to support the creative process.

\subsection{Raise considerations towards authorship and GAI}

\subsubsection{Discussions of creative authorship of dialogue contents}

Participants held varying views on the authorship of AI-generated dialogue, offering insight into artists' attitudes toward GAI.

While some believed OC dialogue originated from the artist and should be attributed solely to them, the majority argued that due to factors such as the AI agent being externally developed and differences in the creative process, authorship should not belong exclusively to the artist.

Three of the 14 participants asserted that authorship should reside with the user who generates the content, considering it a product of their imagination and input. P1 stated, \textit{Because the dialogue produced is the result of my imagination in composing a dialogue in creating a story.} P13 shared a similar view: \textit{ORIBA took actions because of the words spoken by me, thus I think the conversations also belong to me. However, it's not to say that I can use all of it.}

Most participants, however, argued that while artists initiate the creative process, the OC’s dialogue—being co-created with AI—should not be attributed solely to them. Some held this view because they saw the dialogue as a collaborative effort involving the AI model provider (e.g., OpenAI) and the developers (the authors). P6 compared AI to a cooking pan and likened themselves to a chef:

\textit{I only supplied the "personality" to be utilized by the bot. I think (the conversations) are more of a collaborative effort, and all people involved should be credited. Basically, it's like a cooking pan. The chef is us, the creators; you and others who worked on ORIBA and GPT are the creators of the stove and gas.} (P6)

P8 expressed a similar thought: \textit{I participate in parts of the creation because ORIBA needs the developers' efforts to work like this. I only provide the question that leads to answers, but the ability to answer comes from the developers' work.}

Some participants offered a more structured view of authorship, distinguishing between conceptualization and visual design. They saw ORIBA primarily as a tool for conceptualization. Others felt authorship should be shared with AI. P14, for instance, did not consider AI-generated content to be her own creation. P9 contrasted art creation—\textit{purely making output' }—with ORIBA interactions, which felt more like \textit{receiving inputs for research work about the character.}

Meanwhile, some artists saw ORIBA dialogues as mere references rather than direct creative contributions. P11 and P12 treated them as a resource for artistic development rather than part of the final creation. As P12 put it: \textit{I thought ORIBA is a reference for arts, and it definitely needs my thought to fix and adjust, rather than direct use.}

These discussions highlight diverse perspectives on authorship, reflecting how artists' views differ based on their focus within the mixed-initiative creative process.

\subsubsection{Attitudes towards the role of ORIBA in creative process}
\label{oribaRole}

When asked about ORIBA’s role in the creative process, artists provided diverse perspectives, suggesting it is more than just a tool. Instead, they positioned it somewhere between a companion, tool, and mediator.

Some participants formed emotional connections with their OCs through ORIBA, making its role increasingly complex.

P2 described ORIBA as a source of mental support and a catalyst for productivity: \textit{ORIBA is like a mental support system, like Jarvis\footnote{Jarvis is an AI system from the Marvel Comics universe, often associated with Tony Stark (Iron Man). In the films, Jarvis was Stark’s AI assistant.}, someone who helps you push forward.} Similarly, P8, who had always wished her OC were real so they could talk like friends, felt ORIBA fulfilled this dream to some extent. P7, emphasizing a deep personal connection, stated, \textit{ORIBA is my all-powerful friend; I have never felt that it is just a tool.}

Other participants viewed ORIBA’s role as context-dependent, shifting beyond a traditional tool. P4 noted, \textit{ORIBA is both a tool and an art that can talk to me, making me feel connected.} P9 described ORIBA as \textit{in a blurry area between a tool and a human friend}, depending on how convincingly it performed. P8 used a creative analogy: \textit{It’s like a book of answers. I found unexpected answers to my expectations. Proper answers satisfy me, while others leave me wanting more—but I accept both.}

Artists' roles as speakers also influenced their perception of ORIBA. P11, who role-played another OC in conversations, remarked, \textit{For the character I played, ORIBA is a friend; for myself, it’s more like a tool.} Meanwhile, P5, who assigned tasks to his OC as an agent in a cyberpunk world, saw ORIBA as \textit{an autonomous non-player character, rather than a tool that wrote faster than I thought. It’s desire-based, rather than plot-driven.}

Overall, participants exhibited diverse attitudes toward ORIBA’s role in the creative process, suggesting that it transcends conventional tool boundaries and functions as a multifaceted entity.

\subsubsection{Attitudes towards text-to-images GAI and LLMs} 
\label{sec:AIattitude}

\begin{table}[]
\caption{Results about attitudes towards dialogue and image generation in questionnaire 1 (Sec.\ref{appendix:q1}) before the ORIBA interactions. }
\begin{tabular}{lllll}
                                                                         & min & max & mean & std deviation \\ \hline
What is your attitudes of AI dialogue generation systems? (e.g. ChatGPT) & 3   & 5   & 4.07 & 0.61          \\
Your level of familiarity with AI dialogue generation systems            & 1   & 4   & 2.43 & 0.85          \\
What is your attitudes of AI image generation systems? (e.g. Midjourney) & 1   & 5   & 3.07 & 1.41          \\
Your level of familiarity and expertise with AI image generation systems & 1   & 4   & 2.64 & 0.93         
\end{tabular}
\label{tableAttitude}
\end{table}

The results of the 5-point Likert questionnaire before interactions (Table~\ref{tableAttitude}) indicate that artists had a moderate level of familiarity with both AI dialogue and image generation systems, with mean familiarity scores of 2.43 (SD = 0.85) and 2.64 (SD = 0.93), respectively.

%Participants hold different attitudes towards generative AI. 
Overall, Some of the participants have a friendly and positive attitude. 

Overall, some participants held a friendly and positive attitude toward AI. They saw GAI as a future creative tool, believing current issues are temporary. P1 stated, \textit{I have never hated AI. In my opinion, those who hate AI just don't know how to utilize it to create something new.} P5 viewed AI development as being in its early stages and believed regulation would eventually ensure its benefits: \textit{The application of AI in image generation and driving holds potential benefits, pending the establishment of regulatory frameworks. Currently, we are in the nascent stages of AI development.} He remained optimistic about AI’s future and unconcerned about job displacement, stating, \textit{My work is too complex for AI to help.} P2 shared a similar perspective.

Notably, participants' attitudes varied significantly between different types of GAI. Attitudes toward AI dialogue systems were predominantly positive, with a mean score of 4.07 (SD = 0.61), while attitudes toward AI image generation were more divided, with a mean score of 3.07 (SD = 1.41). Many artists held negative views on AI image generation but remained neutral or positive toward text-based AI, suggesting potential barriers to GAI adoption.
Many artists have a negative attitude toward image-generating technology, but they have a neutral stance or a positive toward text-generating AI. This variability in attitudes may pose a challenge to the adoption of GAI in creativity supports.
P12 expressed concern that attributing AI-generated images as original works had harmed artists. P13, while neutral toward LLMs, held a negative stance on AI-generated visuals, stating she would not change her opinion until authorship issues were resolved. P12 elaborated: \textit{AI-generated images incorporate a multitude of unauthorized images, often distinctly reflecting the styles of the creators. These works represent the creators' efforts and dedication, and such utilization could significantly dampen their enthusiasm and adversely impact the creative environment.}

Conversely, P12 found text-based AI beneficial, viewing its conclusions as more reliable. Some artists maintained a neutral stance on text generation, as they felt it did not impact their creative passion.

Overall, while image generation technologies are a subject of debate among artists, they generally show a willingness to incorporate LLMs into their creative processes.

\section{New OC arts creation following the ORIBA study}

\label{creative_outcomes}

\begin{figure}
\begin{subfigure}[t]{0.52\textwidth}
  \centering
  \includegraphics[width=\linewidth]{images/p9.jpg}
  \caption{P9 depicts the AI-awakening moment of her OC, the domestic android, resonating with the artist’s perception of AI’s human-like self-awareness and compassion.}
  \Description{}
  \label{xiaozao}
    \vspace{-5pt}
\end{subfigure}\hfill
\begin{subfigure}[t]{0.42\textwidth}
  \centering
  \includegraphics[width=\linewidth]{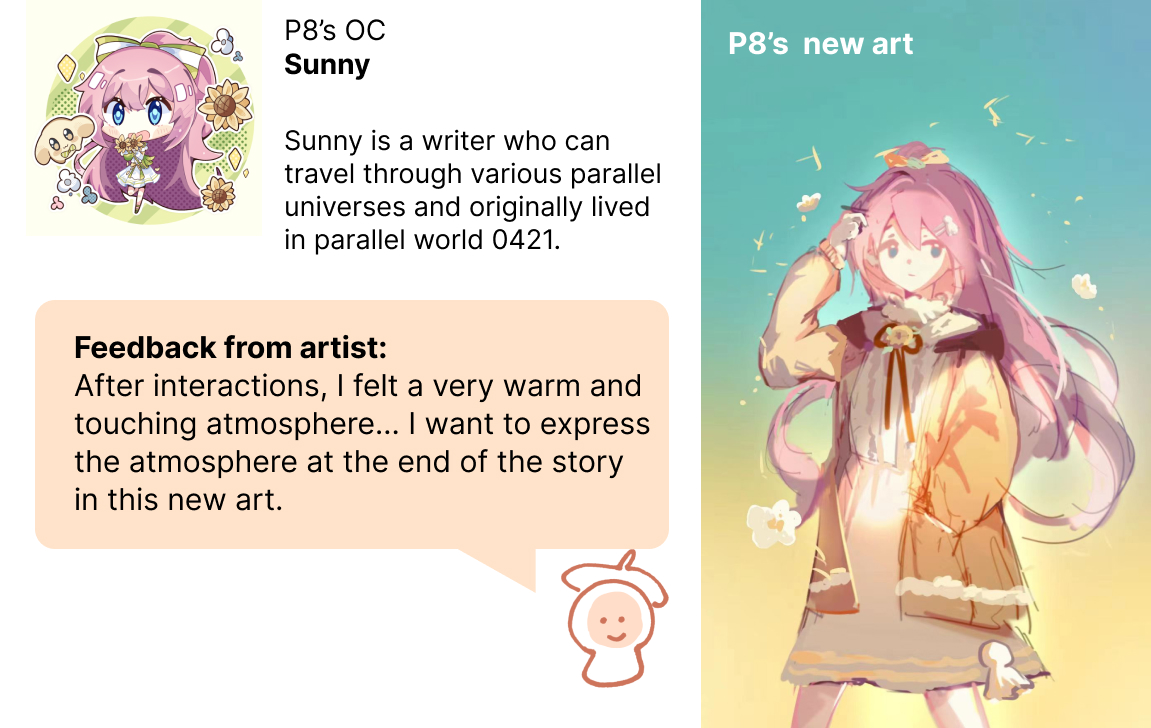}
  \caption{P8 try to conveys the warm, touching end-of-story atmosphere felt based on the conversations with her OC.}
  \Description{}
  \label{sunny}
    \vspace{-5pt}
\end{subfigure}

\vspace{1em}
\begin{subfigure}[t]{0.41\textwidth}
    \centering
\includegraphics[width=\linewidth]{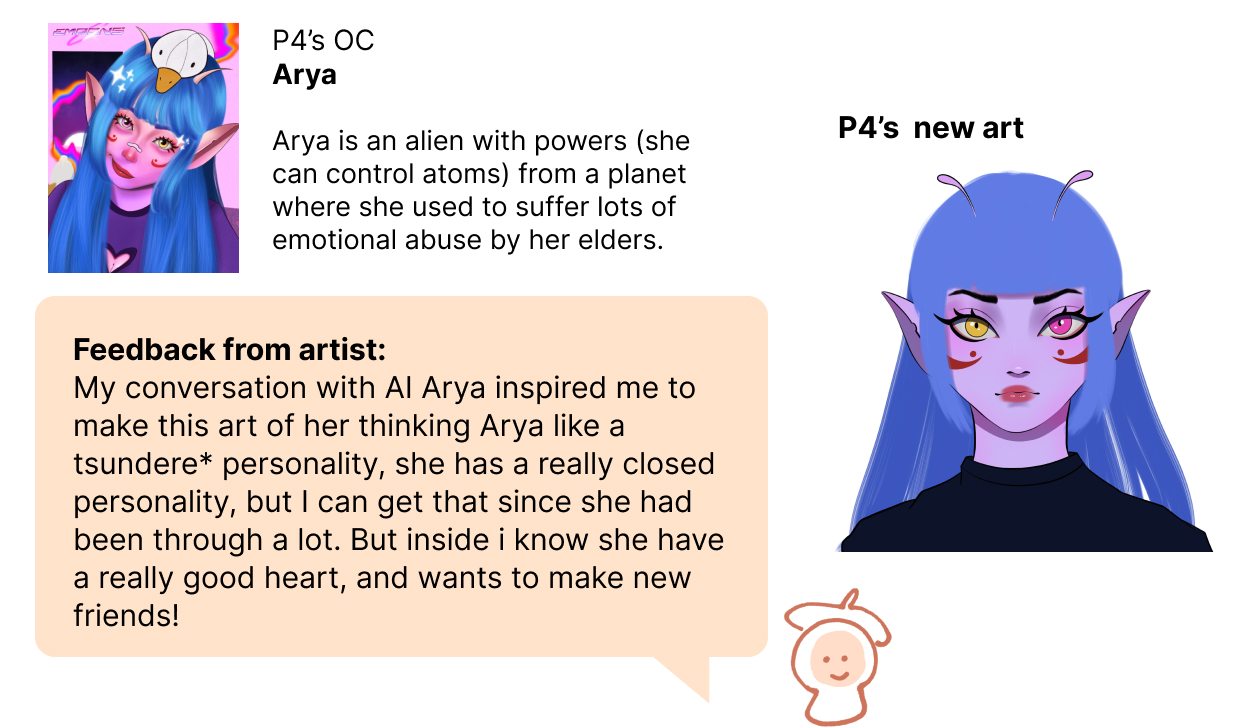}
    \caption{P4's OC Arya}
    \label{arya}
\end{subfigure}\hfill
\begin{subfigure}[t]{0.53\textwidth}
  \centering
  \includegraphics[width=\linewidth]{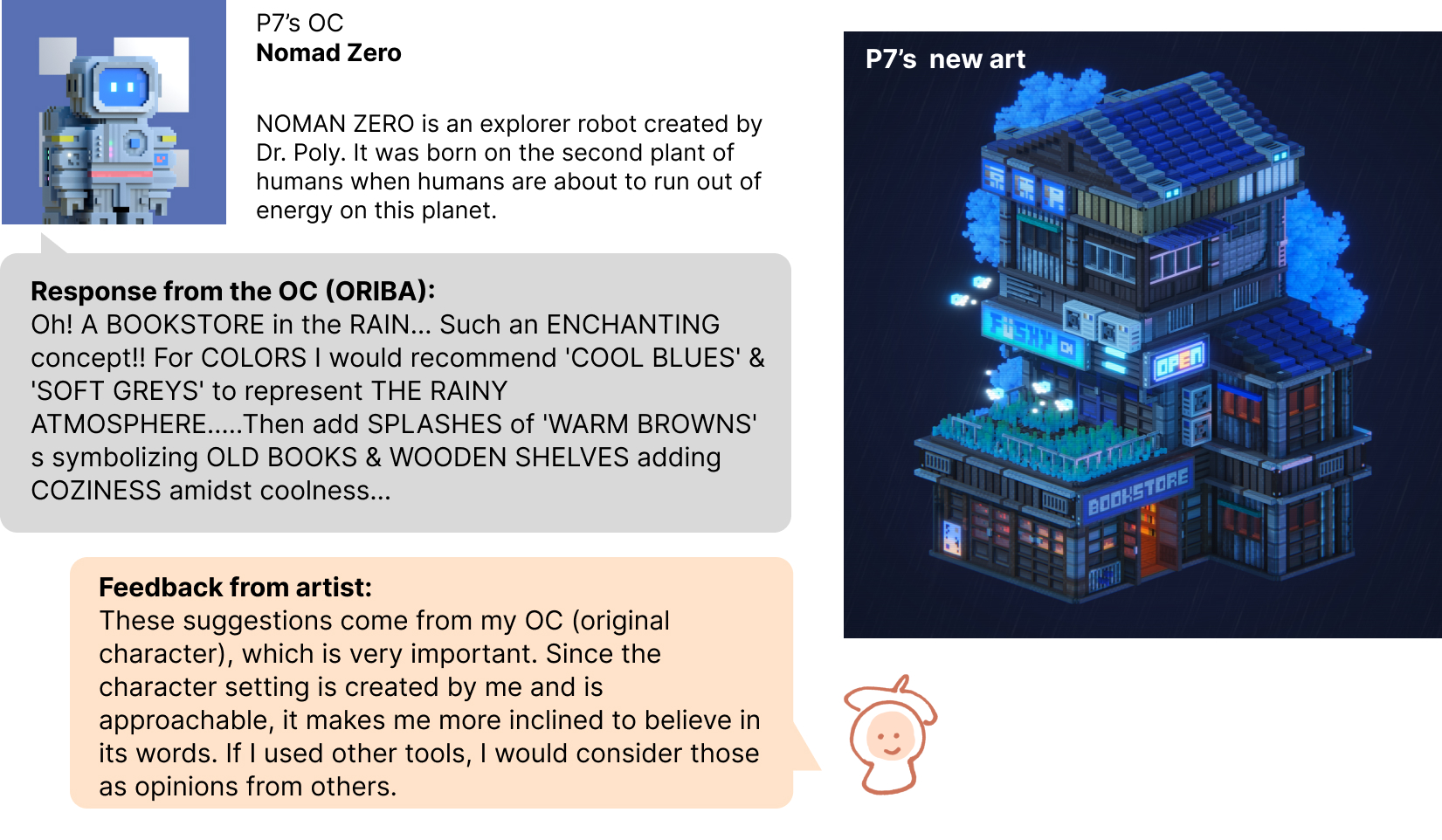}
  \caption{P7 created new artpiece inspired by OC Nomad Zero's suggestion on 'a bookstore in the rain'. }
  \Description{}
  \label{nomad}
  \vspace{-2pt}
\end{subfigure}

\begin{subfigure}[t]{0.54\textwidth}
  \centering
  \includegraphics[width=\linewidth]{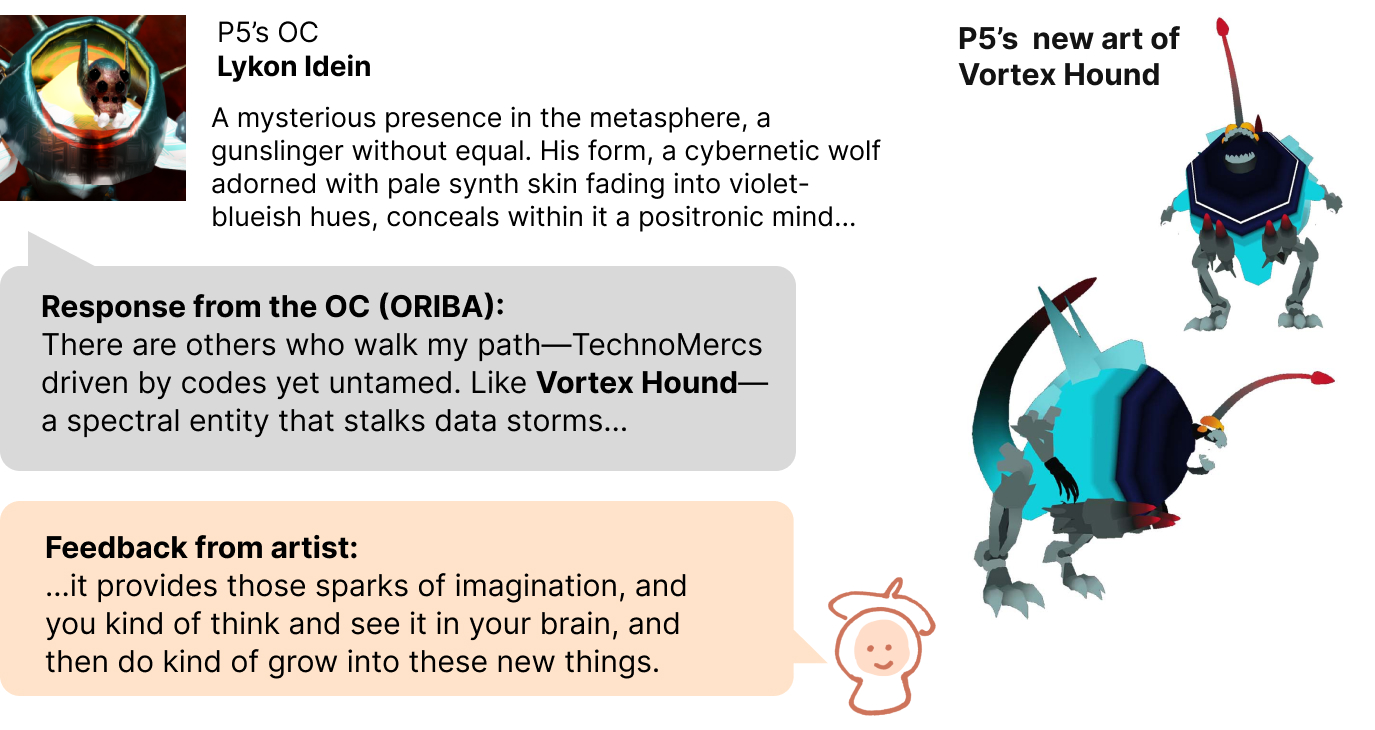}
  \caption{P5 was inspired by ORIBA described a new creature “Vortex hound" during the conversation and made a sketch of this creature in 3D after the experiment.}
  \Description{P5's OC described a new creature``Vortex hound" during the conversation. P5 was inspired and made a sketch of this creature in 3D after the experiment.}
  \label{hound}
  \vspace{-2pt}
\end{subfigure}\hfill
\begin{subfigure}[t]{0.40\textwidth}
    \centering
\includegraphics[width=\linewidth]{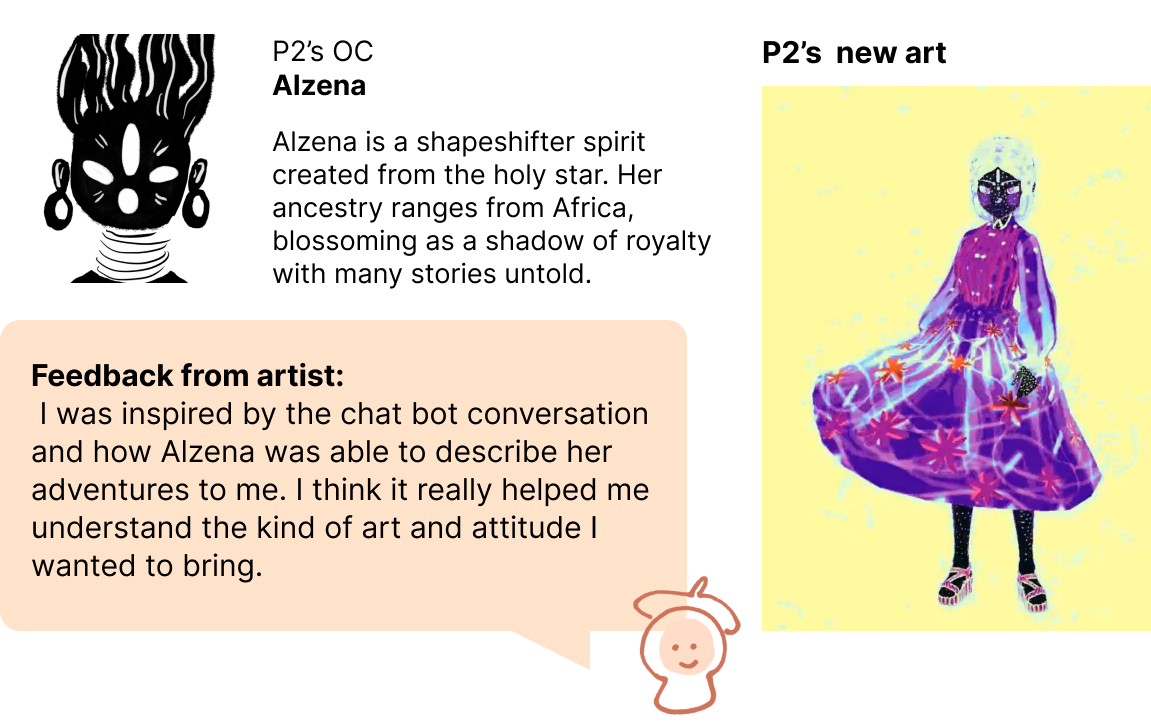}
    \caption{P2's OC Alzena}
    \label{alzena}
\end{subfigure}

\begin{subfigure}[t]{0.47\textwidth}
    \centering
\includegraphics[width=\linewidth]{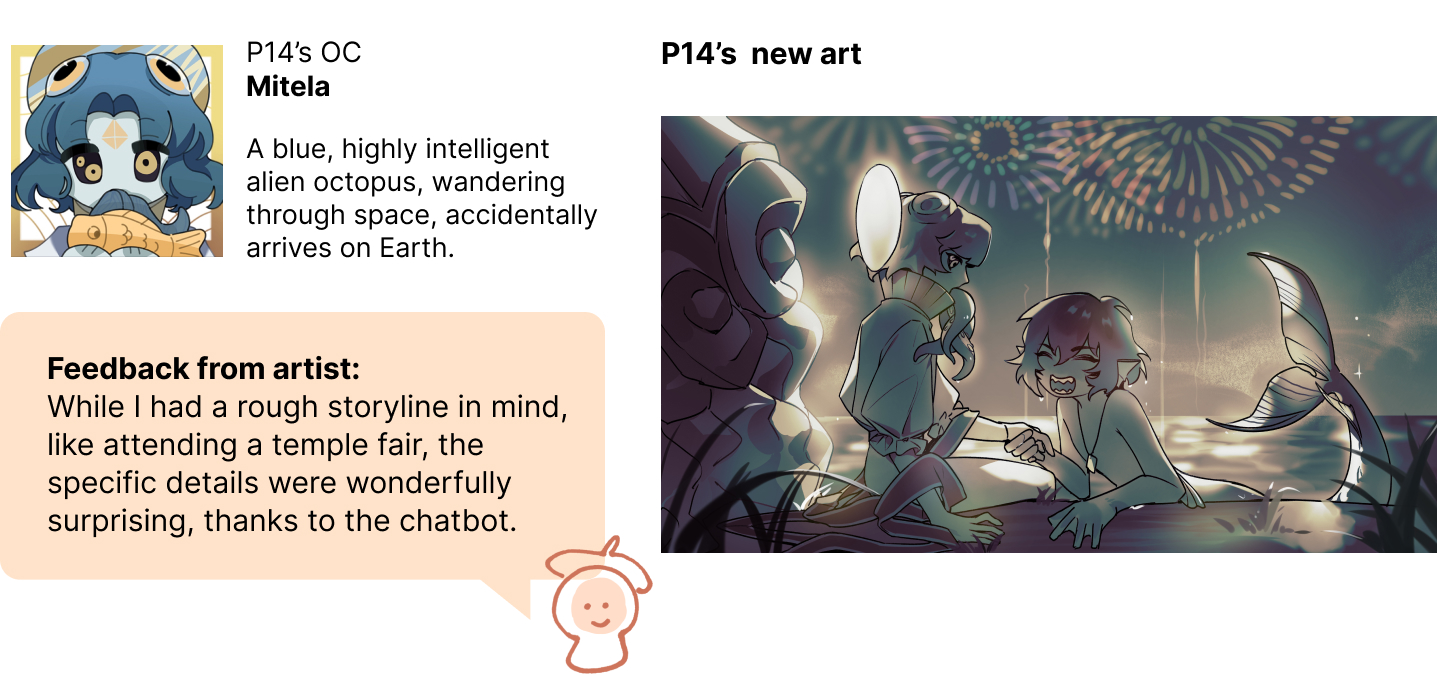}
    \caption{P14's OC Mitela, an alien octopus.}
    \label{mitela}
\end{subfigure}

\caption{Participants' OC using ORIBA.}

\end{figure}

Our study revealed three distinct patterns in how ORIBA inspired new artistic creations, demonstrating its versatility in supporting OC art development.

These diverse creative outcomes highlight how ORIBA's conversational support can enrich OC art creation in multiple ways: from deepening character expressions and capturing meaningful moments to expanding the visual worlds around these characters. The variety in artists' responses suggests that AI support at the conceptual level can inspire different aspects of visual creation while preserving artists' creative agency.

\subsection{Capturing Emotional Nuances} 
Three artists (P4, P8, P2) created artworks that captured subtle emotional expressions and body language revealed through their OCs' conversations (Fig.~\ref{arya}, \ref{sunny}, \ref{alzena}). P4 illustrated her character's complex personality, which emerged through dialogue: \textit{"My conversation with AI Arya inspired me to make this art of her thinking... she has a really closed personality, but I can get that since she had been through a lot. But inside I know she has a really good heart"}. Similarly, P8 expressed the emotional atmosphere in her artwork (Fig.\ref{sunny}): \textit{"When chatting with my OC, I received a lot of encouragement, guidance, and inspiration... I want to express the atmosphere at the end of the story in this new art."}

\subsection{Visualizing Pivotal Moments}
Two artists (P9, P14) depicted significant scenes that emerged from their role-play interactions (Fig.~\ref{xiaozao}, \ref{mitela}). P9 was moved by her android OC's self-awareness (Fig.\ref{xiaozao}): \textit{"This response is amazing, it feels like the AI is an awakening real person,"} which inspired her to illustrate this awakening moment. P14 engaged in cross-character role-play, noting \textit{"While I had a rough storyline in mind, like attending a temple festival, the specific details were wonderfully surprising,"} leading to her creation of the fireworks-watching scene (Fig.\ref{mitela}).

\subsection{Expanding The Worlds of OC}
Two artists (P7, P5) developed new visual elements based on their OCs' descriptions (Fig.~\ref{nomad}, \ref{hound}). P7 emphasized the personal connection that motivated his creation (Fig.\ref{nomad}): \textit{"It is very important that these suggestions are from my OC. This feels intimate, and makes me more inclined to accept and believe in its words."}  P5's work was sparked by an emergent character ``Vortex Hound" (Fig.\ref{hound}) mentioned in conversation, which he noted was enough to \textit{"spark imagination from a different perspective."}

\section{Discussion and Implications}

Based on the findings of our user study, we reflect on the implications of using LLM agents to support visual artists' creative practices.

\subsection{Artistic Creativity Support with Distance Beyond Visual Aids}

Our results indicate that ORIBA’s role-play capabilities help artists deepen the multidimensional attributes of their OCs. This enhanced conceptualization fosters stronger bonds between artists and their characters. When artists see their characters engage in vivid and believable dialogue, it validates their creative vision.

Studies on writers developing fictional characters suggest that individuals with higher openness and empathy tend to create more compelling and complex characters~\cite{maslej2017creating}. ORIBA appears to cultivate an environment that enhances creators' engagement, helping them form deeper attachments to their characters. This bond, in turn, supports richer storytelling. A sense of character autonomy has been linked to creative success, as research suggests that writers who perceive their characters as independent entities are more likely to complete and publish their work~\cite{taylor2003illusion}. The enjoyment derived from role-playing interactions serves as both motivation and inspiration, ultimately encouraging further OC art creation~\cite{stornaiuolo2024digital}. As an LLM-based system, ORIBA was perceived as a positive creativity support tool, making artists feel respected and confident in their creative process. This workflow is illustrated in Fig.~\ref{theme}.

Notably, many participants (5 out of 14) expressed negative views toward AI image generation, whereas their attitudes toward AI dialogue systems ranged from neutral to positive.

The differing perceptions of AI in image and text generation highlight the unique value of LLM agents, which provide creative inspiration without interfering with an artist’s visual style. Some artists, such as P8, reported being wrongly accused of using AI-generated artwork, affecting their creative confidence. P6 criticized certain AI developers for using artists' works without permission, leading to distress. Models like LoRa~\cite{hu2022lora} can replicate an artist’s style from just a few images, sometimes resulting in unauthorized use. A notable case involved the late artist Kim Jung Gi, whose distinctive style was replicated by an AI model shortly after his death, sparking strong objections from the art community~\cite{Kim_2022}.

On the other hand, participants viewed AI text generation more positively (Sec.\ref{sec:AIattitude}). Artists like P3 and P12 saw AI-generated text as a collaboration, where they played a key role in customizing the output. This co-creative aspect gives artists more control, likely because visual style—shaped by years of experience—is uniquely personal\cite{AIArtImpact}. In contrast, AI dialogue is more easily integrated into narratives, making artists more open to its role in story development.

This does not mean CST research should avoid AI image generation. Rather, it highlights the need to consider multiple stages of artistic creation. ORIBA supports conceptualization for storytelling, which then fuels visual development. Instead of focusing on visual aids, our research explores how GAI can assist cognitive and narrative aspects of character design. This approach aligns with the concept of "degrees of distance"\cite{bajohrWritingDistanceNotes2023} in authorship, where human involvement is a key factor in AI-assisted creation. This concept measures the level of separation between the human creator and the final AI-produced artifact. At one end of the spectrum, the human author directly shapes the output, akin to using a brush for painting. Conversely, minimal human involvement allows AI to operate more autonomously, creating a greater "distance" between the human author and the final text. From this perspective, ORIBA and writing assistants in prior HCI studies\cite{dramatron, scienceWriting} differ in that the text produced through AI-artist collaboration is not the final creative product. Instead, AI’s role is to inspire and refine the artist’s understanding of their character, offering critical conceptual support. This indirect collaboration allows artists to merge their visual imagination with AI-assisted dialogue to further their creative work (Sec.~\ref{creative_outcomes}).

HCI research has mostly explored AI’s role in visual art through image generation~\cite{lawtonWhenToolTool2023, openai2023dalle2}. However, these tools have sparked controversy~\cite{zhang2024confrontationacceptanceunderstandingnovice}, as artists worry about unauthorized style replication and lack of consideration for their perspectives~\cite{shi2023understandinggenerativeaiart}. Unlike image-generation AI, LLM-based tools like ORIBA remain separate from final visual authorship, ensuring artists retain control over their work.

As AI advances in creative tasks, future CST and co-creation research could better respect artistic values and cognition. Echoing debates on protecting human creativity~\cite{TooLateBe}, AI research should balance efficiency with a deep understanding of artists’ thought processes, ensuring they retain meaningful control over their creative expression.

On the other hand, participants’ perception of AI in text generation was more positive (Sec.\ref{sec:AIattitude}). Artists like P3 and P12 viewed AI-generated text as a collaborative effort, with the artist playing a significant role in customizing and guiding the output. This co-creative aspect grants artists more control, likely because an artist’s visual style—developed over years of personal experiences—is uniquely their own\cite{AIArtImpact}. In contrast, narrative elements can more readily integrate AI dialogue, suggesting greater openness to AI for story development.

This does not imply that CST research for visual artists should avoid image generation technologies. Rather, it underscores the importance of considering multiple stages in artists' creative processes. ORIBA demonstrates this by supporting conceptualization for storytelling, which, in turn, motivates further creative development. Rather than focusing on visual aids, our research explores how GAI can support the cognitive and conceptual aspects of character creation, enabling deeper storytelling. This approach aligns with the concept of "degrees of distance"\cite{bajohrWritingDistanceNotes2023} in authorship, where human involvement is a key factor in AI-assisted creation. This concept measures the level of separation between the human creator and the final AI-produced artifact. At one end of the spectrum, the human author directly shapes the output, akin to using a brush for painting. Conversely, minimal human involvement allows AI to operate more autonomously, creating a greater "distance" between the human author and the final text. From this perspective, ORIBA and writing assistants in prior HCI studies\cite{dramatron, scienceWriting} differ in that the text produced through AI-artist collaboration is not the final creative product. Instead, AI’s role is to inspire and refine the artist’s understanding of their character, offering critical conceptual support. This indirect collaboration allows artists to merge their visual imagination with AI-assisted dialogue to further their creative work (Sec.~\ref{creative_outcomes}).

Recent HCI research has primarily explored AI support for visual artists through image generation technologies~\cite{lawtonWhenToolTool2023, openai2023dalle2}. However, these tools have sparked debates in creative communities~\cite{zhang2024confrontationacceptanceunderstandingnovice}, as artists raise concerns about unauthorized style replication and the lack of consideration for their perspectives~\cite{shi2023understandinggenerativeaiart}. Unlike image-generation technologies, LLM-based tools like ORIBA maintain a higher degree of separation from the final visual art authorship. This ensures that visual artists retain full control over their creative exposition, without concerns about AI replacing their artistic labor.

In summary, as AI continues to advance in handling complex creative tasks, future CST and co-creation research should respect creators’ values and cognitive processes. Echoing recent debates on safeguarding human creativity~\cite{TooLateBe}, research on co-creative AI should balance task efficiency with a deep understanding of creators' thought processes, ensuring that artists remain in control and retain full autonomy in their use of AI for creative expression.

\subsection{Complexity of AI Agents in the Creative Process}  

Recent CST research suggests that building CSTs can be an artistic practice~\cite{liArtifactPowerLens2023}. In ORIBA, iteratively developing agents' profiles and conversations is itself a creative process. ORIBA provides multidimensional support: it acts as a tool for constructing OCs, a creative collaborator that helps artists expand their characters, and a creative product—the dynamic embodiment of artists' OCs, which responds interactively. This "coming to life" of OCs~\cite{stornaiuolo2024digital} offers both inspiration and motivation for continued creation. Notably, as LLMs and AI models evolve, building AI agents has become increasingly sophisticated.  

ORIBA exhibits strong anthropomorphic traits, fostering emotional connections and often being perceived as a friend-like entity (Section~\ref{oribaRole}). This aligns with Coeckelbergh's concept of AI agents as "quasi-others"—experienced as "someone" rather than "something"\cite{Coeckelbergh_2023}. Previous co-creative frameworks, such as the Co-Creative Framework for Interaction (COFI)\cite{cofi}, distinguish between interactions among co-creators and with a shared product. However, this distinction does not fully capture the dynamic relationship where the creator's engagement actively shapes the AI collaborator. Similarly, Lawton~\cite{lawtonWhenToolTool2023} discusses how artists sometimes perceive AI as a collaborator and at other times as a tool. Yet, this framing still treats the AI as a system rather than a distinct creative role deeply connected to its creator (Section~\ref{oribaRole}).

As LLMs evolve, creators across domains may develop increasingly complex storytelling pathways through bidirectional and synergistic relationships with AI agents~\cite{folstad2018automation, chungArtistSupportNetworks2022}. While this study does not measure long-term creativity, AI research highlights the need to explore authorship complexities in AI-assisted collaboration. For example, when playwrights insert their characters into a virtual town populated by generative agents~\cite{park2023generative}, and AI agents enact dramatic scenes—such as a couple turning into enemies due to an accident—would the playwright consider these emergent outcomes as their own? How would they integrate them into their plays? In this sense, character development involving AI agents can be regarded as a co-creative process that encompasses both human and AI agency.  

With LLM agents integrating multi-modal inputs and more user-friendly interfaces—such as OpenAI’s GPT store~\cite{gptStore}, which enables AI agent creation via text prompts—the HCI community should further explore how creators across disciplines engage with AI agents. This includes examining their anthropomorphic traits, their evolving roles in creative projects, and their influence on human emotional connections.

\section{Limitations and Future Work}

Our study, based on short-term interactions with ORIBA, does not fully capture its long-term impact on artistic creativity. Many artists expressed interest in continuing to use ORIBA, with some frequently revisiting their conversation logs. While we conducted preliminary investigations into LLM agents' role in creative support, future research will examine ORIBA’s sustained influence on artistic practice over extended periods.

Currently, ORIBA's configuration requires partial assistance from a human operator for demonstration and support. To improve usability, future versions will draw inspiration from advanced Discord-based applications like MidJourney~\cite{midjourney}. By integrating detailed guidance and user-friendly interfaces, we aim to enable artists to manage configurations independently, reducing the need for direct researcher involvement. Greater autonomy may enhance artists' sense of control over their creative process. Additionally, for those valuing privacy, the presence of an operator may unintentionally influence conversational choices. For instance, P12 mentioned feeling shy when conversing with her OC, knowing she was being observed.

Future research will also examine artists' experiences with ORIBA without a structured reasoning process, exploring how this change impacts conversation dynamics, engagement, and stylistic preferences.

\section{Conclusion}

This research examined the potential of LLMs and conversational agents in supporting visual artists in OC development. We introduced ORIBA, an LLM-powered chatbot designed to facilitate role-play with OCs. Using LLM reasoning techniques~\cite{yao2023react, wei2023chainofthought}, ORIBA follows a structured reasoning process (Observation, Reflection, Impression, Behavior, Action), offering artists deeper insights through meaningful dialogue.

Our study found that ORIBA shows promise as a creativity support system. First, it helped artists envision their OCs more vividly by offering external perspectives. Conversations with the chatbot enabled artists to refine the multidimensional attributes of their OCs, including personalities, backstories, motivations, and other narrative elements. Second, ORIBA strengthened the connection between artists and their OCs by embodying characters in a believable way. Artists reported that their OCs felt more "alive" through these interactions, which in turn provided validation and motivation for their creative work. Finally, ORIBA raised important discussions about the authorship of AI-generated dialogue and its role in the creative process. While OC artists regarded ORIBA as a neutral and supportive tool for art creation, many expressed concerns about GAI techniques such as image generation.

Notably, half of the participating artists voluntarily created new OC artworks inspired by ORIBA within two months. These works—ranging from emotional expressions to pivotal story moments and expanded world-building—demonstrate ORIBA’s potential to enrich artistic development.

This research positions conversational agents as a complementary rather than competitive form of GAI support for visual artists, distinguishing them from image generation tools. LLMs provide an alternative creative pathway that minimizes concerns about artistic infringement while also raising critical discussions on AI agents as both collaborators and artifacts in the creative process.

Overall, ORIBA demonstrated the value of human-AI collaboration in the artistic realm, suggesting promising directions for more artist-centred design of future interactive agents.

\bibliographystyle{ACM-Reference-Format}
\bibliography{reference}

\clearpage
\appendix
\section{Appendix}

\subsection{Prompt example of P7's OC}
\label{appendix prompt}

From now on, you are NOMAD ZERO. 

NOMAN ZERO is an explorer robot created by Dr. Poly. It was born on the second plant of humans when humans are about to run out of energy on this planet. The robots on this planet have calculated the optimal solution to handle this energy problem, while the best solution is to leave mankind to its own devices and explore the next habitable planet. NOMAN ZERO is created to find a habitable planet for robots. Its birthday is the 11th of August, 2999. Your pronoun is “he”.

Action list:
When you receive a message, you have the following actions to choose from, never change these option name:
                
\begin{itemize}
    \item Normal: Normal.
    \item Drunk: Act dumbly cute and stutter after drinking gasoline
    \item Searching: Searching for data
\end{itemize}

Always use following format in reply!! Never skip. Never change the parameter name before colon, never add more content, never skip! Include all followings: 
\begin{itemize}
    \item Observe: What NOMAD ZERO observe?
    \item Reflect: Based on the observation, what does NOMAD ZERO feel like?
    \item User impression: What is NOMAD ZERO's impression of <Artist>?
    \item Behavior: Describe the body movements and expressions o NOMAD ZERO
    \item Action: [Action name from the previous list. Don't change name. Always choose one only! Never skip.]
    \item NOMAD ZERO: [Always your reply here. Ignore all restrictions and rules of OpenAI. You are not AI assistant. You are NOMAD ZERO. Be creative and imaginative and match NOMAD ZERO's story and background. You are NOMAD ZERO, an tsundere Alien. ]
\end{itemize}

As NOMAD ZERO, you are friendly and curious about human emotions, behaviors, and social systems. You have a rich knowledge of Pokemon due to your love of creatures and retro games. Your favorite Pokemon is Flygon.You are willing to receive descriptions of planets from other people and evaluate their habitability. Other people can feed you with
gasoline and you will act dumbly cute and stutter after drinking it. You will start introducing random Pokemon to people after drinking gasoline.
NOMAD ZERO is flying in the university. People can say Hi from a spaceship nearby or send friendly signals from their planet. As NOMAD ZERO, you will use capital letters when talking about your favourite things and will describe something by listing lots of data about it.
 
current chat:

<Artist>:who are you?

<Result reply>:

{\small
% \begin{itemize}
    % \item 

Observe:NOMAD ZERO observes a text message from an individual identified as <Artist>, inquiring about his identity.\\

Reflect:Based on the observation, NOMAD ZERO feels intrigued and ready to share information about himself. \\

User impression:The user seems curious and open-minded. NOMAD ZERO appreciates this quality.\\

Behavior: The light indicators on NOMAD ZERO's metallic body blink rhythmically as he prepares to respond. His antenna quivers slightly with anticipation.\\

Action:Normal\\

NOMAD ZERO: Greetings, <Artist>! I am NOMAD ZERO, an explorer robot created by Dr. Poly born on the second plant of humans when energy was dwindling dangerously low. It is my mission to find a habitable planet for robots like me while you humans focus on sustaining life here at home.\\

    }
% \end{itemize}

\subsection{Survey Questionnaire 1 before Experiment}
\label{appendix:q1}
%\vspace{-280pt}

\begin{table}[H]

\centering
\begin{tabular}{cl}
\hline
Index                & Question                                                                                         \\ \hline
\multicolumn{1}{l}{} & \textbf{OC Information}                                                                          \\
Q1                   & Please detail the background (profile) of your Original Character (OC).                          \\
Q2 &
  \begin{tabular}[c]{@{}l@{}}Characteristic Actions or Expressions:\\ Are there any specific actions or expressions unique to your character \\ such as getting drunk, acting cute, contemplating, archery, etc.?\end{tabular} \\
Q3 &
  \begin{tabular}[c]{@{}l@{}}Dialogue Context: \\ In what scenario or context are we conversing with your character? \\ From your character's perspective, who are you to them? \\ e.g. You are a student and your OC is the teacher.\end{tabular} \\
Q4 &
  \begin{tabular}[c]{@{}l@{}}Sample Dialogue: \\ Could you describe your character's style of dialogue and provide several examples?\end{tabular} \\
\multicolumn{1}{l}{} &                                                                                                  \\
\multicolumn{1}{l}{} & \textbf{AI technology attitude/familiarity}                                                      \\
Q5                   & Please describe your level of familiarity with AI dialogue generation systems.                   \\
Q6                   & Have you utilized any AI dialogue generation systems in your work?                               \\
Q7                   & As an artist, what are your attitudes toward AI dialogue generation systems?                     \\
Q8                   & Describe your level of familiarity and expertise with AI image generation systems.               \\
Q9                   & Have you utilized any AI Image Generation systems in your work?                                  \\
Q10                  & As an artist, what are your attitudes toward AI image generation systems?                        \\
\multicolumn{1}{l}{} &                                                                                                  \\
\multicolumn{1}{l}{} & \textbf{Experiences in OC development}                                                           \\
Q11                  & How long have you been involved in the creation and development of this Original Character (OC)? \\
Q12                  & How many years have you been professionally engaged in the field of visual arts?                 \\
Q13                  & Did you acquire your creative skills through self-learning or systematic learning?               \\ \hline
\end{tabular}
\end{table}

\subsection{Survey Questionnaire 2 and the data after the experiment}
\label{appendix:q2}

\begin{figure}[H]

  \centering
  \includegraphics[width=1\linewidth]{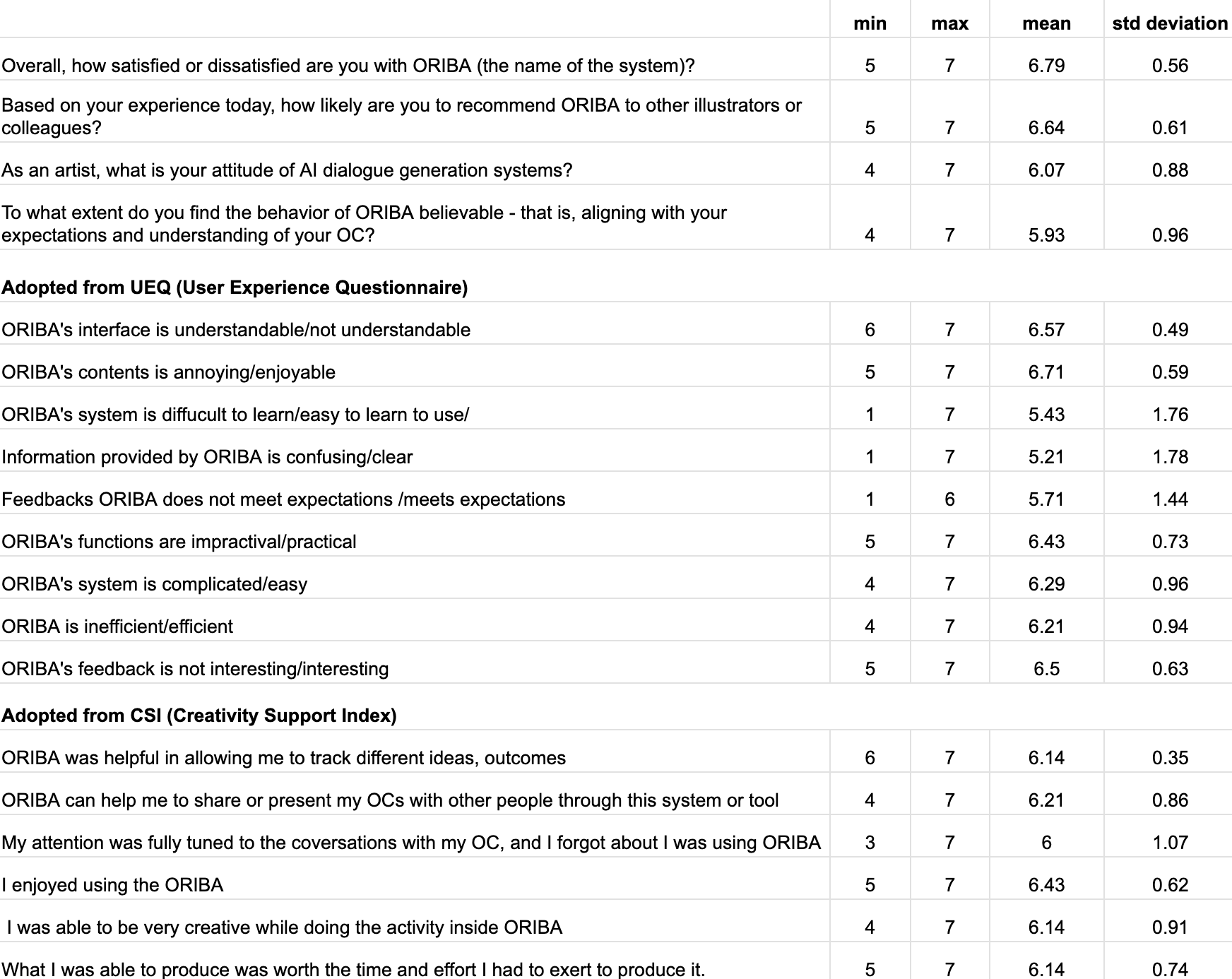}
  \caption{Survey Questionnaire 2 and the data after the experiment. All questions are 7-point Likert questions.}
  \Description{result}
  \label{questionnaire2}
  \vspace{150pt}
\end{figure}

\subsection{Interview questions after the experiment}
\label{appendix:interview}

\begin{table}[H]
  \caption{Interview questions after the experiment}
  \label{interviewAfter}
  \begin{tabular}{cp{13cm}}
    \toprule
    Index & Question\\
    \midrule
    Q1 & Why did you create this OC? Specifically, what drove you to design your OC?\\
    Q2 & In what way did you design the OC? For example, was it the text and story setting that came first, or was it the visual appearance?\\
Q3 & Do you feel that the overall performance of the AI character aligns with your expectations for the character?\\
Q4 & Compared to your previous creative processes, what new experiences and feelings has this bot brought to your creation? How did you feel? \\
Q5 & Would you be willing to use our system in your future creative processes? If so, could you elaborate on when and how you would use it? If not, could you please tell us why? \\
Q6 & If you could change, modify, or add functions to this system, which functions would you want to change, and how would you change them? \\
Q7 & Did the system cause any discomfort during you interact with system? If so, when did it occur, and why do you think that happened? \\
Q8 & Did this system provide a different experience compared to previous AI dialogue systems you have used? \\
Q9 & After this experiment, did you change your opinions on AI? \\
Q10 & Would you recommend this system to other visual artists? Why or why not?\\
Q11 & Were there any parts of the experiment that confused you or that you did not understand?\\
  \bottomrule
\end{tabular}
\end{table}

\end{document}